\documentclass[12pt]{article}
\usepackage{amsmath, amssymb, graphics}
\usepackage{amsbsy}
\usepackage{amsfonts}
\usepackage{amsmath}
\usepackage{graphicx}
\newcommand{\mathsym}[1]{{}}

\def\ci{\cite}

\def \S {{\rm S}}

\def \td {\tilde}

\def \N {{\mathcal N}}

\def \Str {{\rm Str}}
\def \m {\mu}

\def \vep {\varepsilon}

\def \bi{\bibitem}
\def \la {\label}

\def\foot{\footnote}

\def \adss {$AdS_5 \times S^5$\ }
\newcommand{\rf}[1]{(\ref{#1})}
\def \ov {\over}

\def\N{{\cal N}}

\def \ha{{1\ov 2}}

\def \del {\partial}

\def \S {{\cal S}}

 \def \bb {\bar \begin{equation}ta}

\def \bi{\bibitem}
\def \la {\label}

\def\foot{\footnote}

\def \adss {$AdS_5 \times S^5$\ }

 \def \p {\phi}

\def \ov {\over}

\def \varpi {{\rm w}}

\def \rt {{\rm t}}
\def \no {\nonumber }

\def \adss {$AdS_5 \times S^5$\ }

\def \N {{\cal N}}

\def \bb {{\bar  \begin{equation}ta}}

  \def \te {\theta}

\def \dpp {\del_+}
\def \dmm {\del_-}
\def \k {\kappa}
\def \ci {\cite}

\def\tr{{\rm tr}}
\def\str{{\rm Str}}

\newcommand{\psr}{\Psi_{{}_R}}
\newcommand{\psl}{\Psi_{{}_L}}

\renewcommand{\tilde}{\widetilde}
\renewcommand{\hat}{\widehat}

\newcommand{\algf}{\Liealg{f}}
\newcommand{\alghf}{\hat\algf}

\newcommand{\algh}{\Liealg{h}}

\newcommand{\Liealg}{\mathfrak}       

\newcommand{\id}{\mathbf{1}}
\renewcommand{\imath}{i}

\newcommand{\binner}[2]{%
  {\langle}\kern-4.15pt{\langle}#1{,}\,#2{\rangle}\kern-4.15pt{\rangle}}

\newcommand{\ffrac}[2]{\raisebox{.5pt}%
  {\footnotesize$\displaystyle\frac{#1}{#2}$}\kern1pt}

\def\const{\mathop\mathrm{const}\nolimits}

\def\id{\protect{{1 \kern-.28em {\rm l}}}}

\def\p{{\partial}}

\makeatletter
\renewcommand\section{\@startsection {section}{1}{\z@}%
                                   {-3.5ex \@plus -1ex \@minus -.2ex}%
                                   {2.3ex \@plus.2ex}%
                                   {\normalfont\large\bfseries}}
\renewcommand\subsection{\@startsection{subsection}{2}{\z@}%
                                   {-3.25ex\@plus -1ex \@minus -.2ex}%
                                   {1.5ex \@plus .2ex}%
                                   {\normalfont\normalsize\bfseries}}
\makeatother

\def\b{{\rm b}} 

\def\STr{{\rm Str}}
\def\Tr{{\rm tr}}

\def \vp {\varphi}

\def \L {\Lambda}
\def \bs {\bigskip}

\numberwithin{equation}{section} \makeatletter
\@addtoreset{equation}{section}

\vspace{-1cm}

  \def \te {\theta}
  \def \vp {\varphi}
  \def \a {\alpha}
  \def \b {\beta}

\newcommand{\pp}{{\parallel}}

\def \A {{\cal A}}

\def \s {\sigma}

\def \vt  {\vartheta}

\def \CS {{\Sigma }}

\def \n {\nu}
\def \P {\Phi}

\newcommand{\be}{\begin{eqnarray}}
\newcommand{\ee}{\end{eqnarray}}

\def \ze {\zeta}

\def \vt {\vartheta}
\def \U {{\cal U}}

\begin{document}

\vspace{ -3cm}
\rightline{Imperial-TP-AT-2009-1}


\begin{center}
\vspace{0.2cm}
{\Large\bf
UV finiteness of Pohlmeyer-reduced form
  \\
\vspace{0.15cm}
 of the  $AdS_5 \times S^5$ superstring theory \\
\vspace{0.2cm}
\vspace{0.4cm}
   }

 \vspace{0.5cm} {R. Roiban$^{a,}$\footnote{radu@phys.psu.edu}
 and A.A. Tseytlin$^{b,}$\footnote{Also at
 Lebedev Institute. 
  tseytlin@imperial.ac.uk
 }}\\
 \vskip 0.13cm

{\em
$^{a}$Department of Physics, The Pennsylvania  State University,\\
University Park, PA 16802 , USA\\
$^{b}$   Blackett Laboratory, Imperial College,
London SW7 2AZ, U.K.     
  }

\end{center}

 \begin{abstract}
  We consider the Pohlmeyer-type reduced theory  found
 by explicitly  solving  the   Virasoro constraints
 in the formulation of \adss superstring in terms of
 supercoset currents. The resulting set of classically equivalent,
 integrable Lagrangian equations of motion has the advantage of
 involving  only a physical number of degrees of freedom
 and yet being 2d Lorentz invariant.
 The corresponding  reduced theory action may be written  as  a
 gauged WZW model coupled to fermions with further bosonic and fermionic
 potential terms. Since the \adss superstring sigma model is
 conformally invariant,  its classical relation to the reduced theory
 may extend to the quantum level only if the latter 
 is, in fact,  UV  finite. This theory is power counting renormalizable
 with the only possible divergences being of potential type. 
 We explicitly verify its  1-loop finiteness and 
 show   that the   2-loop
 divergences are, in general,  scheme dependent  and 
 vanish  in  dimensional reduction scheme.
We expect that
 the reduced  theory is finite to all  orders in the loop expansion.

\end{abstract}
\thispagestyle{empty}
\setcounter{page}{0}
\newpage

\setcounter{footnote}{0}

\def \rb {{\rm b}}

\def \G {\Gamma} 

\def \eps {\epsilon}
\def \diag {{\rm diag}}
\def \TT  {{\rm T}} 
\def \rt {{\rm t}} 
\def \bs {\bigskip}
\def \km {{\mathbf{\mu}}} 

\def \ph {\phi}

\def \psr {\Psi_{_R}} \def \psl {\Psi_{_L}}
\def \g {\gamma} 

 \def \by  {\times }
 \def \lol {{_{L}}}
  \def \rol {{_{R}}}
\def \cP {{\cal P}}
\def \da {\dagger}
\def \S {\Sigma} 
\def \ha {{\textstyle {1 \ov 2}} }
\def \de {\delta}
\def \bp {\begin{pmatrix}}  \def \ep {\end{pmatrix}}
\def \af   {\alghf}
\def \X {{\cal X}}
\def \tg {{\tilde \g}}
\def \tb {{\tilde \b }}
\def \td {\tilde}
\def \JJ {{\rm J}}
\def \RT  {reduced theory\ }
\newpage
\renewcommand{\theequation}{1.\arabic{equation}}
 \setcounter{equation}{0}

\setcounter{equation}{0} \setcounter{footnote}{0}
\setcounter{section}{0}

\def \bea {\be}
\def \eea {\ee}

\def \p {\phi}

\def \U {U} 

\section{Introduction}

Recent remarkable progress in understanding the spectrum of states
with large quantum numbers in \adss string theory or dual $\N=4$ SYM
theory was achieved via interplay of various perturbative data from
gauge theory and string theory linked together by the assumption of exact
integrability. It remains an outstanding problem to derive the
corresponding asymptotic Bethe ansatz equations directly from
first principles -- from quantum superstring theory. That would be
facilitated if the corresponding integrable \adss sigma model admitted
a formulation in terms of elementary excitations with two-dimensional Lorentz
covariant S-matrix. Such a formulation may also make more
straightforward 
the generalization of the asymptotic Bethe Ansatz to the case
when both strings and dual operators have finite length, i.e. to the
case of closed strings on the cylinder $R_t \times S^1$.
%

With this motivation in mind here we shall continue the study of the
Pohlmeyer-reduced 
\ci{pol}
formulation of gauge-fixed \adss superstring
\ci{gt,ms,gt2}. This theory (which we shall refer to  as the
``reduced theory'') is a generalized sine-Gordon or non-abelian
Toda type two-dimensional Lorentz-invariant sigma model which 
is closely related
to the original Green-Schwarz (GS) superstring  sigma model
\ci{mt}. It is constructed by writing the GS superstring  
equations of motion in
terms of the components of the 
$PSU(2,2|4)\ov SO(1,4) \times SO(5)$ supercoset current, 
fixing the conformal and $\kappa$-symmetry gauges and then
reconstructing the action that reproduces the equation of motion for
the remaining physical number of degrees of freedom.

 While the resulting \RT is classically equivalent to the original \adss GS superstring 
  (and, in particular, it is also  classically integrable) it is a
 priori unclear if the corresponding quantum theories should be
 closely related. In general, 
 the classical Pohlmeyer reduction assumes two-dimensional conformal
 invariance but for sigma models with target spaces involving $S^n$ or
 $AdS_n$ factors (and no bosonic WZ couplings) that symmetry may hold also 
 at the quantum level only in very exceptional cases like the  \adss GS superstring.
 The minimal consistency requirement for the conjecture that the
 classical equivalence between the GS superstring  and the \RT may extend to the
 quantum level is then the finiteness of the \RT -- the cancellation of
 the UV divergences  in world-sheet perturbation
 theory. This means the absence of any new dynamically generated scale
 in addition to the classical mass parameter in the potential
 introduced in the process of fixing the classical conformal
 diffeomorphism symmetry  (this procedure  spontaneously breaks the
 underlying conformal symmetry of the GS superstring  in conformal gauge while
 preserving two-dimensional Lorentz invariance). 
 
 Our aim below will be to  demonstrate the cancellation of the 1-loop and 2-loop
 divergences in the \RT  which also gives a strong indication 
 of all-loop finiteness. 

\

Let us first briefly  discuss  what is known about the  \adss superstring
theory. The classical theory \ci{mt}  generalizes the \adss
bosonic sigma model to the presence of GS fermions incorporating
self-dual 5-form coupling. 
The potential importance of integrability 
of this model  (motivated by  the known integrability of its bosonic part) 
was recognized early on  \ci{mtt,wad}; the classical intergrability   was 
 proved  in the full theory including fermions in 
 \ci{bpr} (see also \ci{beis,roibs};  for a review see
\ci{afr}).
 Given the
global symmetry, uniqueness of the (2-derivative) action and analogy
with WZW theory the action is expected to be UV finite to all orders
\ci{mt} and that was directly verified at 1-loop \ci{dgt,ft1} and
2-loop \ci{rt} orders. The classical integrability appears to extend to
the quantum level as is effectively verified by the matching  of the
1-loop
\ci{ft1} and 2-loop \ci{rt} corrections to spinning string energies
to  the strong-coupling predictions of the asymptotic Bethe Ansatz (see, e.g.,
\ci{btz} and \ci{benna}).\foot{Quantum integrability was also argued
for in the closely related pure spinor formulation of \adss
superstring \ci{berk,mss,pull}.}
 
The GS action has a well-known peculiarity in that to carry out its 
perturbative expansion it is necessary to choose a non-trivial
background  for  the closed string coordinates  and expand around it. 
The background introduces a fiducial mass scale (spontaneously breaking
two-dimensional conformal invariance) and also spontaneously breaks
the two-dimensional Lorentz invariance at the level of interaction
terms in the action.
That happens, for example, when  one expands near a null geodesic or uses a
version of light-cone (l.c.) gauge in \adss \ci{mtt,call,af}.  While this
step is a natural one when computing quantum superstring corrections
to specific string states, it is a complication in general
considerations (e.g., in computing the underlying 
factorized
S-matrix). In particular, the l.c. gauge fixed
\adss GS superstring  action   has a complicated interaction 
structure making the direct computation of the corresponding
magnon-type or BMN excitation S-matrix problematic beyond the tree
level \ci{roibs}.  Another complication is that when formally expanded
near a particular background the GS action is not power-counting
renormalizable \ci{poly,rt} and one is to rely on a judicious choice
of regularization (and measure) to verify the cancellation of the UV
divergences.

Remarkably, these problems are absent in the quantum theory as defined
in terms of the supercoset current variables, i.e. defined by the \RT action \ci{gt}. The
corresponding fermionic kinetic terms have standard two-dimensional Dirac form
and thus the two-dimensional Lorentz covariant fermionic propagators are
defined without independently of  a bosonic string  coordinate  background. Moreover,
 the \RT action is
power counting renormalizable and relatively straightforward to
quantize, as its structure is similar to that of two-dimensional
supersymmetric gauged $G/H$ WZW model supplemented with  a bosonic potential
and a ``Yukawa'' interaction term.\foot{As we shall see, the ``Yukawa'' 
interaction is effectively 
responsible for the UV finiteness of the ``gWZW+potential'' model.} 
%

The quadratic part of the \RT action has the same form as that of the
GS superstring expanded near the BMN vacuum, i.e. as the GS action in
maximally-supersymmetric plane wave background in the l.c. gauge
\ci{mets,bmn}:
%
eight two-dimensional scalars together with eight two-dimensional
Majorana fermions, all with equal mass $\mu$.  The interaction terms
differ, but one may hope that there exists a certain transformation
relating the corresponding S-matrices.\foot{These are expected to have
closely related symmetries: $PSU(2|2) \times PSU(2|2)$ in the GS
superstring case
\ci{beisert,af} and $SU(2)\times SU(2)\times SU(2)\times SU(2)$ in the \RT
case \ci{gt} -- the latter is formally the same as the 
bosonic part of the former but their precise relation 
needs to be clarified further.
} 
Since both the \adss superstring and the corresponding \RT are
expected to be conformal theories, the parameter $\mu$ should be the
only scale on which the quantum S-matrices should depend.  While the
S-matrix corresponding the BMN vacuum is not two-dimensional
Lorentz invariant, the one appearing in the \RT should be Lorentz
invariant (i.e. the 4-point scattering matrix should depend only on
the difference of the two rapidities). This puts the reduced theory
into the same class of integrable theories as the solvable $O(n)$ sigma
models.
 
 This motivates the  
 study of the \RT at the
 quantum level even regardless 
 its relation to the quantum GS superstring  theory: it
 appears to be   a  remarkable  finite
 integrable model with several unique features.
 
 \

 Below in section 2 we shall start with a review of the \RT action 
  using an explicit parametrization of the fermionic variables
 and clarifying on the way several important features of this
 theory. As was already mentioned, the construction of Pohlmeyer-reduced
 theory (see \ci{gt} and also 
  \ci{mira} and references therein) involves several steps:
  
 (i) start with the GS equations (and the Maurer-Cartan equations)
 written in terms of the components of the ${\hat F \ov G}=
 {PSU(2,2|4)\ov SO(1,4) \times SO(5)}$ supercoset current;
 
 (ii) solve the conformal gauge
 constraints introducing a new set of field variables directly
 (algebraically) related to the currents, fixing the residual
 conformal diffeomorpisms and $\kappa$-symmetry gauge in the process;
 
 (iii) reconstruct an action for the remaining field equations in
 terms of the new (physical)  variables. 
 
 \noindent 
 The resulting \RT action defines a massive integrable
 two-dimensional field theory. Its construction thus involves a
 non-local map between the original coset coordinate fields and
 current variables that preserves the integrable structure and allows
 the reconstruction of the classical solutions of the GS superstring
 action from classical solutions of \RT action, i.e. the solitonic
 solutions in the two models are in direct correspondence.\foot{This
 correspondence was used in \ci{hofm}.}  

The bosonic fields of the \RT are $g \in G= Sp(2,2) \times Sp(4) \subset
PSU(2,2|4)$  and  the two-dimensional gauge field $A_\mu$ taking values 
in the algebra of
$H= SU(2)\times SU(2)\times SU(2)\times SU(2)
\subset G$. In addition, there are fermionic fields $\psr,\psl$  
(directly related to fermionic currents of the GS superstring) which are
two-dimensional Majorana spinors with the { {standard}} kinetic terms
transforming under both $Sp(2,2)$ and $ Sp(4)$ and thus linking together
the two sets of bosons (corresponding effectively to the ``transverse'' string
fluctuations in $AdS_5$ and $S^5$).\foot{
This model is kind of ``hybrid'' of  a WZW model based on a supercoset (where
fermions are in ``off-diagonal'' blocks of a supermatrix field but
have non-unitary second-derivative kinetic terms) and  a two-dimensional
supersymmetric version of a $G/H$ gWZW model where fermions have the 
standard first-order kinetic terms but take values in the coset part of
the algebra of the group $G$.}  In the special case when \adss is replaced
by $AdS_2 \times S^2$ the corresponding \RT is equivalent \ci{gt} to the ${\cal
N} =2$ super sine-Gordon model (there $H$ is trivial).

At the level of the equations of motion of 
the \RT 
it is possible to fix the $A_\mu=0$ gauge;  
the equations then  become equivalent to a fermionic
generalization of  non-abelian Toda equations.  The linearization
of the equations of motion in the gauge $A_\m=0$ around the trivial
vacuum $g=\id$ gives 8+8 bosonic and fermionic degrees of freedom with
mass $\mu$ and suggests that the symmetry of resulting {relativistic}
S-matrix should be $H=[SU(2)]^4$.

 The potential term   is  multiplied by the
 ``built-in'' classical  scale parameter $\mu$  
which is a remnant of gauge-fixing the conformal
 diffeomorphisms at the classical  level.  
Consistency then requires that the reduced theory 
be also UV finite, i.e. while a priori   the 
 $\m$-dependent terms in the \RT action may renormalize, 
 the 
fermions  should {{cancel}} the bosonic renormalization.

  This is indeed what happens in the $AdS_2 \times S^2$ case (i.e. in
  the ${\cal N}=2$ super sine-Gordon model).  As we shall see in section 4
  below, this is also true in the general \adss case: we shall
  demonstrate the cancellation of UV divergences at the 1-loop and
  2-loop orders in the natural dimensional reduction regularization
  scheme.\foot{The same scheme was used in \ci{rt} where the 2-loop
  finiteness of the \adss GS superstring was verified.}  We believe
  that similar cancellations should extend to all orders in
  perturbation theory.  Then the theory
  is UV finite and $\mu$ remains an arbitrary conformal symmetry gauge
  fixing parameter at the quantum level.  The cancellation of
  divergences 
  is presumably related to a hidden symmetry that should have its
  origin in $\kappa$-symmetry of the original GS action that relates
  the coefficients of the ``kinetic'' and the WZ terms in the action
  (which, under the reduction, become the potential and the Yukawa
  terms in the reduced action).
  
There are several conceptual issues that remain to be clarified before
one would  be able to claim that the quantum \RT is indeed directly relevant
for solving the quantum \adss superstring theory.  These include the precise mapping
between observables and conserved charges (cf. \ci{gt2}) and
understanding the relation between massive S-matrices computed by
expanding near the respective vacua. The ultimate motivation for the study
of the \RT is the hope that it may be more straightforward to   define as a quantum 
integrable theory
and thus easier to solve
 than the original \adss GS superstring  model. To demonstrate this remains a
 program for the future.

\renewcommand{\theequation}{2.\arabic{equation}}
 \setcounter{equation}{0}

\section{Reduced theory  for \adss superstring}

In this section  we shall  review the structure of the reduced theory action.

Our starting point is the \adss superstring action \ci{mt} written in
terms of currents for the supercoset\foot{The bosonic part of the
$PSU(2,2|4)$ group is $SU(2,2) \times SU(4)$ or $ SO(2,4) \times
SO(6)$ and an equivalent form of the subgroup is $SO(1,4) \times
SO(5)$.} $${\hat F\ov G}= { PSU(2,2|4)\ov Sp(2,2) \times Sp(4) } $$
The currents take values in the superalgebra $\hat \algf= psu(2,2|4)$
which is a quotient of $su(2,2|4)$ by elements proportional to unit
matrix.

Let us first discuss the explicit parametrization of the corresponding
supermatrices.

\subsection{Supercoset  parametrization, currents  and gauge fixing}

An  element  of $su(2,2|4)$  can be written as an $8 \times 8$  matrix 
\be \la{xx}
M=\bp
A&X\\
 X^\da \S &B
\ep \ , \ \ \  \  \str\ M= \tr A-\tr B=0 \ , \ \ \  \  
 A \in u(2,2)\ , \ \ \ \ \  B \in u(4)  . \ee
Let us  define the $4 \times 4$
matrices $\S$ and $K$ 
(we follow the notation of \ci{ar,gt,afr}; $I$
 denotes a unit matrix of an appropriate dimension)
\begin{equation}\la{kk} 
 \S=\bp
I &0\\
0&-I \ep,\quad K=\bp
J&0\\
0&J \ep ,  \ \  \ J=  \bp
0&-1\\
1&0  \ep  , \ \ \    \  \  [\S, K]=0, \ \  \S^2 = I,  \  \ K^2=-I  
 \end{equation}
The superalgebra $su(2,2|4)$ admits a $Z_4$ automorphism \ci{bersh},
i.e.  its elements can be split into four orthogonal subspaces
$\af_0\oplus \af_1 \oplus \af_2\oplus \af_3$, with $[\af_i, \af_j] =
\af_{i+j\ (mod\ 4)}$ in the following way:
\be \la{bb}
M_{0,2} = \bp
A_{0,2}&0\\
0&B_{0,2}
\ep \ , \ \ \    A_{0,2}= \ha ( A  \pm   K A^t K)\ , \ \ \  \ \ 
B_{0,2} =\ha ( B   \pm  K B^t K ), 
\ee
\be 
M_{1,3} = \bp
0& X_{1,3}    \\
 X^\da_{1,3}\S  &0
\ep \ , \ \ \    X_{1,3} = \ha [ X  \pm i   K ( X^\dagger \S)^t K]
= \ha ( X  \pm i  \S  K   X^* K )
 . \la{pro}
\ee
Here $A_0 \in sp(4), \ B_0 \in sp(4)$, i.e. $M_0$ belongs to $sp(2,2)
\oplus sp(4)$, while $M_2$ is in the bosonic part of the coset
subspace of the algebra.  $M_1$ and $M_3$ are expressed in terms of
the real and imaginary parts of the complex $4 \times 4$ matrix $X$.
This split is a ``reality decomposition'' implemented by the
projectors applied to X:
\be \la{poj}
X_{1,3}  =  \cP_{{_\pm}} X  \equiv  \ha ( X \pm  i \S K X^* K)  \ , \ \ \ \ \ \ \  \ \
\ 
\ \ \cP^2_{{_\pm}} =\cP_{{_\pm}} \ .   
\ee
Thus the elements from $\alghf_1$ and $\alghf_3$ should satisfy the
following conditions
\be \la{req}
 X^*_1  = -  i \S K X_1 K \ ,\ \ \ \ \ \ \ \ 
 X^*_3  =   i \S K X_3 K \ ,  
\ee
 which can be solved  explicitly  in terms of  $4 \by 4$ matrices $\X_{1,3}$
 with independent {\it real}  Grassmann elements 
\be 
 X_1 =  \X_1 + i \S K \X_1 K \ , \ \ \ \ \ \  
 X_3 =  \X_3 - i \S K \X_3 K \ .  
\la{res} 
\ee
 
 \
 
The \adss GS action \ci{mt,bersh,roisig} is constructed by starting
with an element $f$ of $\hat F= PSU(2,2|4)$, defining the current
$\JJ= f^{-1} d f$ and then splitting the current according to the $Z_4$
decomposition of $\hat \algf$  ($\m,\n=(0,1)$)
\begin{equation}
\JJ_\m=f^{-1}\del_\m f= \A_\m  +Q_{1\m}+P_\m +Q_{2\m} \,,\qquad \qquad 
\A \in \alghf_0,\quad Q_1\in \alghf_1,\quad P\in\alghf_2,\quad Q_2\in \alghf_3\ .
\end{equation}
Here $\A_\m$  belongs to  the algebra of the 
subgroup $G$ defining  the $\hat{F}/G$ coset,  
i.e. $G=  Sp(2,2) \times Sp(4)$  (isomorphic to $SO(1,4)\times SO(5)$), 
$P$ is in the bosonic coset  (i.e.  \adss)  component, 
and  $Q_1,\ Q_2$ are the fermionic  currents.  The  Lagrangian 
 in conformal gauge is  
 \be \la{al}
L= \str\left[P_+P_-  + \ha (Q_{1+}Q_{2-}-Q_{1-}Q_{2+})\right]\,,
\ee
which should be supplemented by the Virasoro (conformal gauge)   constraints 
\be  \str  (P_+P_+) =0 \ , \ \ \ \ \  \str  (P_-P_-) =0  \ . \ee
As already reviewed in the introduction, the idea behind the
construction of the reduced action \ci{gt,ms} is to express the
corresponding equations in terms of currents only, solve the conformal
conformal gauge constraints algebraically introducing a new set of
field variables directly related to the currents, then 
choose a
$\kappa$-symmetry gauge and finally reconstruct the action
corresponding to the resulting field equations in terms of current
variables. This construction implies the classical equivalence of the
original and ``reduced'' sets of equations; in particular, the reduced
theory is also integrable \ci{gt}.

The Virasoro constraints   can be solved  by fixing a special 
 $G$-gauge and residual conformal diffeomorphism gauge such that
 \foot{In general, 
 we may introduce  two  different parameters $\mu_+$ and $\mu_-$  in $P_+$ and $P_-$;
 the resulting expression for the reduced action will then  be obtained by replacing 
 $\mu \to \sqrt{ \mu_+ \mu_-}$.} 
 \begin{equation}
  \la{ret} 
P_+=\ \km \ T\ ,\qquad \qquad P_-=
\ \km\  g^{-1}Tg\,,\ \ \ \ \ \ \ \ \   \km=\const \ . 
\end{equation}
Here $\mu$ is an arbitrary scale parameter (the scale  corresponding
to fixing the residual conformal diffeomorphisms, similar to $p^+$ in 
light-cone gauge) and $T$ is a special constant matrix  chosen in 
\ci{gt} to be\foot{The choice of normalization of
 $T$ is of course arbitrary and can be changed by  rescaling  $\mu$.} 
\be \la{ip}
T={ i \ov 2}  \bp
\S& 0    \\
0 &\S
\ep \ , \ \ \ \ \ \ \ \  \  \ \ \ \ \ 
T^2 = - {1 \ov 4} I ,\ \  \ \  \ \ \str\ T^2 =0   \ .  \ee 
Here $ \S$ is defined in \rf{kk}   and  we also introduced   a  new bosonic field variable $g$  which belongs 
 to   $G= Sp(2,2) \times Sp(4)$, i.e. to the 
 subgroup whose Lie algebra is $\alghf_0$. 
 
 Having chosen $T$, we may define a subgroup $H$ in $G$ that commutes
 with $T$, \ $[T, h]=0, \ h \in H$: in the present case we get $H=
 SU(2) \times SU(2) \times SU(2) \times SU(2)$.\foot{Note that there
 is a natural arbitrariness in the choice of $g$ in eq.~(\ref{ret})
 since $P_-$ is invariant under $g\to hg$ if $h\in H $; that implies
 an additional $H$ gauge invariance of the resulting equations of
 motion for $g$.}
%
%
Using the gauge freedom and the equations of motion one can choose
$g\in G $ and $A_+, A_-$ taking values in the algebra $\algh=su(2)
\oplus su(2)\oplus su(2)\oplus su(2)$ of $H$ 
and defined by 
\be \la{ko} A_+\equiv g
\A_+ g^{-1} + \del_+g\ g^{-1} \,, \ \ \ \ \ \ \ \ \ A_- \equiv
(\A_-)_\algh \ee
 as the new independent bosonic variables \ci{bps,gt}.
   
  Next, one can    impose a  partial $\kappa$-symmetry gauge
\be
\la{iu} Q_{1-}=0 \ ,\ \ \ \ \ \ \ \    \ \ \ Q_{2+}=0  \ ,  
\ee
and then define the new independent fermionic variables
\begin{equation}\la{psa}
\Psi_1=Q_{1+}\,,\qquad \qquad \Psi_2=g Q_{2-}g^{-1}\ . 
\end{equation} 
Similarly to $Q_{1+}$ and $Q_{2-}$, the new variables $\Psi_1$ and
$\Psi_2$ belong to $\alghf_1$ and $\alghf_3$, respectively.  Indeed, 
the adjoint action of $g \in G$ separately maps the subspaces
$\af_1$ and $\af_3$ into themselves 
%
%
since the algebra of $G$ is $\af_0$ and according to the $Z_4$
decomposition $[\alghf_i,\alghf_j]=\alghf_{i+j \ mod \ 4}$.
i.e. $[\alghf_0,\alghf_3]=\alghf_{3}.$
Note also that $\Psi_1$ and $\Psi_2$ are completely independent being
related to different components of the fermionic current.

The residual $\k$-symmetry can be fixed by further 
restricting $\Psi_{1,2}$  by   demanding 
that they  anticommute with $T$,  $\{\Psi_{1,2}, T\}=0$.
Namely, we may introduce the projector from $\Psi_{1,2}$ to $\Psi^\pp_{1,2}$
\bea   \la{coon}
&& \Psi^\pp \equiv  \Pi   \Psi =  \ha (\Psi  +  4 T \Psi T) \ , \ \ \ \ \  \Pi^2 = \Pi
 \ ,\ \\ 
&& \ 
\Psi^\pp T = - T \Psi^\pp \  ,   \ \ \  \ \ \ \ \ \ 
[T, [T, \Psi^\pp]] = - \Psi^\pp  \ , \ \ \ \ \ \ \ 
\Psi^\pp = [T, \hat \Psi] , \ \ \ \  \hat \Psi = -2 T \Psi   \ .  \la{hhh} \eea 
Note that since according to \rf{kk} 
 $[\S,K]=0$  the projector $\Pi$   commutes with the 
``reality condition''  projectors  $\cP_{{_\pm}} $ in \rf{poj}, 
so that it   can be imposed in addition 
to the constraints \rf{req} or \rf{res}.

The $Z_2$ decomposition implied by   $\Pi$ can be represented  explicitly  
as follows: 
\be \la{pio}
\Psi= \bp
0&  X    \\
 X^\da\S  &0
\ep, \ \ \  \ \  \   X= X^\pp + X^\perp, \ \ \ \ \ 
  X^\pp = - \S X^\pp \S, \ \ \ \  X^\perp =  \S X^\perp \S    \ . \ee
Writing  $X$ in terms  of $2 \times 2$ blocks  and  using \rf{kk} 
 we get 
\be \la{spl} 
X\equiv  \bp
\a &  \b    \\
\g  &   \de   \ep, \ \ \  \ \ \ \ 
X^\pp = \bp
0 &  \b    \\
\g  &  0    \ep,\ \ \ \ \ \  X^\perp = \bp
\a &  0    \\
0  &  \de   \ep  \ . 
\ee
We may  then define the new fermionic variables   as \ci{gt} 
\begin{equation}\la{psis}
 \Psi_{_R}={\textstyle \frac{1}{\sqrt{\km}}} \Psi_1^\pp\,,
 \qquad\qquad \Psi_{{_L}}={\textstyle\frac{1}{\sqrt{\km}}}
 \Psi_2^\pp\,, 
\end{equation}
so that $\Psi_{_R}$ and $\Psi_{_L}$    are  expressed in terms of 
``off-diagonal''  matrices $X_\rol$ and $X_\lol$  as $X^\pp$  in
 \rf{spl}.
 The ``reality'' constraints  \rf{res}  on $\Psi_{_R}\in \af_1$ and 
 $\Psi_{_L}\in \af_3$  in \rf{res}   then    imply that 
 the corresponding $2\by2$ blocks  are expressed  
  in terms of {\it real}  Grassmann $2 \by2 $ matrices $\xi$ and $\eta$
 ($J^2= -I$, see \rf{kk}) 
\be \la{kpq}
\b_{_{R,L}} =  \xi_{_{R,L}}   \pm   iJ \xi_{_{R,L}}  J \  , \ \ \  \ \ \ \ \ \ \ \ \ 
\g_{_{R,L}} = \eta_{_{R,L}}  \mp   iJ \eta{_{R}} J \ .
\ee
Explicitly, in terms of $2 \by 2$ blocks
\bea \la{exl}
&&\Psi_{_R}= \bp 0& 0 & 0&  \xi_{_{R}}   +  iJ \xi_{_{R}}  J \\
              0& 0 &  \eta_{_{R}}  -   iJ \eta_{_{R}} J &  0\\
              0&   -   \eta^t_{_{R}}  -   iJ \eta^t_{_{R}} J    &  0  &  0\\
             \xi^t_{_{R}}   -  iJ \xi^t_{_{R}}  J   & 0    &  0  &  0\ep  \ , \\
\la{exz}
&&\Psi_{_L}= \bp 0& 0 & 0&  \xi_{_{L}}   -  iJ \xi_{_{L}}  J \\
              0& 0 &  \eta_{_{L}}  +   iJ \eta_{_{L}} J &  0\\
              0&   -   \eta^t_{_{L}}  +   iJ \eta^t_{_{L}} J    &  0  &  0\\
             \xi^t_{_{L}}   +  iJ \xi^t_{_{L}}  J   & 0    &  0  &  0
\ep  \ .\eea
Thus each of $\Psi_{_R}$ and $\Psi_{_L}$  are parametrized by 
$2 \by 4=8$ independent real Grassmann  variables. Note that the change
$R \to  L$  is equivalent to $ i \to -i$, i.e. 
\be   \Psi_{_R} ( \xi_R , \eta_R)   = \Psi_{_L}^* (\xi_L \to \xi_R, \eta_L \to \eta_R)   \ . \la{coom}   \ee

\subsection{Lagrangian of the reduced theory}

The \RT Lagrangian that reproduces the classical equations of the
reduced theory (obtained from first-order equations corresponding to
the GS Lagrangian \rf{al}) is given by the left-right symmetrically
gauged WZW model for $${G\ov H}= { Sp(2,2) \ov SU(2) \times SU(2) }
\times { Sp(4) \ov SU(2) \times SU(2) } $$
 supplemented by  the following  integrable bosonic 
   potential and  the  fermionic terms \ci{gt}:
\bea 
\la{lag}
L_{tot}= L_B + L_F = &&L_{\rm gWZW} (g, A ) +\km^2\, \str (g^{-1}Tg T ) \no \\
&& +\ \str \left ( \Psi_{_L} T D_+\Psi_{_L}  +  \Psi_{_R} T D_-\Psi_{_R}
+\ \km\, g^{-1}\Psi_{_L}g \Psi_{_R}\right)\,.
\eea 
Here all fields are represented by $8 \by 8$
supermatrices (so that $\str$ in bosonic terms means the difference of
traces of the $su(2,2)$ and $su(4)$ parts.  The covariant derivative
is $D_\pm \Psi = \del_\pm \Psi + [A_\pm, \Psi], \ A_\pm \in
\algh$. Given that $[T, h]=0$, $h \in H$, the Lagrangian $L_{tot}$ is
invariant under $H$ gauge transformations
\be  
g' = h^{-1} g h , \  \ \  \ \ 
A_\pm'=  h^{-1} A_\pm h  + h^{-1} \del_\pm h , \ \ \  \ \ 
\Psi'_{_{L,R}} = h^{-1} \Psi_{_{L,R}} h   \ . 
\ee 
The $\m$-dependent terms in \rf{lag} are essentially the original GS
Lagrangian after the substitution of \rf{ret},\rf{iu}, \rf{psa} and
\rf{psis}; one may conjecture that $L_{\rm gWZW} (g, A )$ plus free
fermionic terms should originate from the change of variables (from
fields to currents) in the original GS string path integral
\ci{gt,gt2}.

Similarly to the original closed string GS action, the reduced theory
action is defined on a 2d cylinder (i.e. the fields are $2 \pi$
periodic in $\s$) and should also have the string tension in front of
it. In discussing UV (short distance) behavior of the theory the
compactness of the $\s$ direction is not relevant; likewise the
masses of fields are also unimportant.
In that discussion we shall therefore formally replace the cylinder
with coordinates $(\tau, \s)$ by a plane and consider the mass terms
as part of the interaction potential.  In that case the parameter
$\mu$ (which, as we shall see will not be renormalized) can be set to
$1$ by rescaling the worldsheet coordinates; we will prefer however
not to do that explicitly.

The dimension of the bosonic target space in \rf{lag} is the same as
the dimension of the $G/H$ coset, i.e. 4+4=8. The fermionic fields
having ``standard'' two-dimensional fermionic kinetic terms are
represented by the $8\times 8$ matrices subject to the two $Z_2$
grading conditions discussed above, so that they are
describing eight left-moving and eight right-moving Grassmann degrees
of freedom.  Remarkably, the reduced action is only quadratic in
fermions, in contrast to original GS action which is at least quartic
in fermions in a generic real $\kappa$-symmetry gauge.

Another  way  of writing the fermionic terms, which
takes into account the constraint $T \Psi_{_{L,R}}= - \Psi_{_{L,R}}
T$, follows from introducing an explicit projector in the fermion
kinetic term, as was done in \ci{gt}: 
 $ \Psi T D \Psi \to    \Psi T \Pi D \Psi$. The resulting action is\foot{The
projectors   in the interaction term may be omitted as they will be implemented 
in perturbation theory  through the fermionic propagator factors.
 Another equivalent way of writing the action is 
 solve the constraint $\{ T, \Psi\}=0$ as $\Psi= [ T, \hat \Psi]$.} 
\be \la{kop}
L_F = \ha  \str \left (  \Psi_{_L} [T,  D_+\Psi_{_L}]  +  \Psi_{_R} [T, D_-\Psi_{_R}]
+\ 2\km\, g^{-1}  \Pi \Psi_{_L}g  \Pi \Psi_{_R}\right)\,. 
\ee
The second ``reality'' constraint \rf{res}
 implied by the $Z_4$ split  may also be implemented  by insertion of the corresponding 
 projectors. 
 
One may  also write the action in terms of the independent
 real Grassmann variables entering  the explicit
 solution \rf{exl},\rf{exz} of the constraints. 
 Using \rf{exl},\rf{exz} fermionic kinetic term  in $L_F$ then
  takes the standard  simple form (upon integration by parts)\foot{Note that (up to a total
  derivative) 
  $ \str ( \Psi T d \Psi)  = - {i \ov 2}  \tr [X^ \da ( dX -  \S dX \S)]$, 
  where we used eq.~(\ref{poj}) and the fact that the  fermionic matrices
  anticommute  under the  ordinary trace.}
 \be 
 \la{ki}
 L_{F0} =\str  ( \Psi_{_L} T \del_+\Psi_{_L}  +  \Psi_{_R} T \del_- \Psi_{_R})
 = 
 -2i \ \tr ( \xi^t_{_L} \del_+\xi_{_L} + \eta^t_{_L} \del_+\eta_{_L} +
 \xi^t_{_R} \del_-\xi_{_R} + \eta^t_{_R} \del_-\eta_{_R})
 \ . 
 \ee
 The gauge connection in $D_\pm$ which belongs to $\algh= 
 su(2) \oplus su(2) \oplus su(2)\oplus su(2)$   can be easily included. 
 If $A=\diag(A_1,A_2,A_3,A_4), \ \ A_i \in su(2)$
 then  we get terms like 
$\tr[\b^\da (A_1\b - \b A_4) - \g^\da (A_2\g - \g A_3)]$.
 Then the action can be rewritten in terms of independent $2 \by2$  
 matrices.\foot{Expanding near the trivial solution $A=0,\ g=1$ 
 the fermionic action then takes the form equivalent  to the 
 quadratic fermionic action in the near - pp-wave or BMN limit in eqs. (5.6),(5.7) 
 in \ci{roibs}.}

 The ``Yukawa'' interaction term  in \rf{lag} can be written in more explicit form by using
 that 
 \be \la{jee}
 g = \bp g^{(1)} & 0\\ 0 & g^{(2)}  \ep \ , \ \ \ \ \ \ g^{(1)} \in Sp(2,2), \ \ \  
  g^{(2)} \in Sp(4)
 \ee 
 \be \la{yuu}
 \str  (g^{-1}  \Psi_{_L} g   \Psi_{_R}) = 
 \tr ( g^{(1)}{}^{-1}  X_{_L}  g^{(2)} X^\dagger_{_R} \S - g^{(2)}{}^{-1} X^\dagger_{_L} \S g^{(1)} X_{_R}  ) \  ,  \ee 
 where  
 \be \la{kyu}
  X_{_{R}} = \bp  0 & \xi_{_{R}}   +  iJ \xi_{_{R}}  J \\
                \eta_{_{R}}  -   iJ \eta_{_{R}} J &  0\ep     \ , 
 \ \ \ \ \ \ \     X_{_{L}} = \bp  0 & \xi_{_{L}}   -  iJ \xi_{_{R}}  J \\
                \eta_{_{L}}  +  iJ \eta_{_{L}} J &  0    \ep   \ . 
\ee 		
 This fermionic interaction term is the only one that that mixes the
 bosonic fields $ g^{(1)} \in Sp(2,2)$ and $ g^{(2)} \in Sp(4)$ of
the reduced models (based on gWZW models for ${ Sp(2,2) \ov SU(2)
\times SU(2) }$ and ${ Sp(4) \ov SU(2) \times SU(2) }$) for the
$AdS_5$ and $S^5$ parts of the original GS coset model.\foot{A similar
term in the original GS action reflects the presence of the RR 5-form
coupling.}  The fermions carry representations of both $Sp(2,2)$ and $
Sp(4)$ and thus intertwine the two bosonic sub-theories.\foot{This
feature resembles more a WZW models based on a supergroup rather than
a supersymmetric extension of WZW model.  At the same time, the
fermions here have first-order kinetic term, so we obtain a kind of
hybrid model. In the special case of $AdS_2 \times S^2$ the resulting
reduced model does have 2d supersymmetry and is equivalent to the 
${\cal N}=2$ supersymmetric extension of the 
sine-Gordon model. In this case $G=
SO(1,1) \times SO(2)$ so the fermions are in the singlet
representation. A less trivial case of the reduced model for $AdS_3
\times S^3$ was worked out explicitly in \ci{gt2}; there the
existence of the 2d supersymmetry in the resulting model is not
obvious and remains an open question.}
%

It is this interaction that is responsible for making the reduced model
UV finite, i.e.  conformally invariant modulo the built-in scale
parameter $\mu$ (which is the remnant of gauge-fixing the conformal
diffeomorphisms at the classical level).

At the level of the equations of motion the $H$ gauge field $A_\pm$
can be gauged away; the result is the 
%
%
following fermionic generalization of the non-abelian Toda equations
\ci{gt} (see also \ci{gomis}) 
\bea \la{tod} 
&&\del_- ( g^{-1} \del_+ g) + \mu^2 [ g^{-1}
T g, T] + \mu [g^{-1} \Psi_{_L} g, \Psi_{_R}]=0 \ , \\ 
&&  \del_- \Psi_{_R} -2  \mu  T (g^{-1} \Psi_{_L} g)^\pp =0 \ , \ \ \ \ \  \del_+
\Psi_{_L} -2  \mu T (g \Psi_{_R} g^{-1} )^\pp =0 \ ,\la{od} \\ 
&& ( g^{-1} \del_+ g -2 T \Psi_{_R}\Psi_{_R} 
)_\algh =0 \ , \ \ \ \ ( g
\del_- g^{-1} - 2 T \Psi_{_L}\Psi_{_L} 
 )_\algh =0 \ , \la{kuk}
\eea 
where the last line follows from the equations for $A_\pm$
and we used that $\Psi_{_L,R}$ anticommute with $T$
(see \rf{coon},\rf{psis})  as well as that $T^2 = -{ 1 \ov 4} I$. 
  
 One may also eliminate the gauge fields from the fermionic terms in
 \rf{lag} as usual in 2 dimensions -- by writing $A_+ = u \del_\pm
 u^{-1} $, \ $A_+ = \bar u \del_\pm \bar u^{-1} $ and performing a
 local rotation of the fermions.\foot{As in the supersymmetric WZW
 model, the corresponding Jacobian may lead to a shift of the
 coefficient of the bosonic term.}  The bosonic gWZW part of the
 Lagrangian written in terms of $h_{{_\pm}}$ becomes $L_{_{\rm WZW}} (
 u^{-1} g \bar u) - L_{_{\rm WZW}} ( u^{-1} \bar u )$ and the
 potential term can also be written in terms of $\td g= u^{-1} g \bar
 u $ since $T$ commutes with $u, \bar u$.
 
 Alternatively, one may fix  an $H$ gauge on $g$ and integrate the fields $A_\pm$ out 
 \ci{gt} leading
 to a bosonic sigma model with 4+4 dimensional target space  coupled to 8 fermions 
 (with quadratic and quartic fermionic terms).\foot{ 
 A disadvantage of this gauge   is that the resulting action does not allow a 
 straightforward
 expansion near the $g=1$ point. For this purpose it seems necessary
to choose a ``intermediate'' gauge, where both $A_\pm$ and $g$ are 
partially fixed.}
%
 
 \

Since the  fermions are transforming in different representation than bosons,
the reduced Lagrangian \rf{lag} is not of a familiar supersymmetric
gWZW theory (deformed by a bosonic potential and Yukawa-type terms)
and thus more difficult to analyze. It is nevertheless a simple
well-defined theory intimately connected to the \adss GS superstring.
%
%
It is therefore of interest to study its 
 quantum properties. Finiteness   of \adss superstring 
 (checked directly  to the two-loop order \ci{rt})  suggests, assuming the relation via the reduction should
 hold beyond the classical level, 
  that this  theory should also be UV finite. In contrast to the  GS  superstring, 
   here it  should
  be much easier to verify the finiteness 
   since the reduced  theory is power counting renormalizable. 
   
   Indeed,  the \RT 
   is obviously UV finite for $\mu=0$ (since gWZW  model coupled to fermions is).  
  Also,  the structure of the $\mu$-dependent interaction terms in \rf{lag}
  is constrained  by symmetries, and it seems  possible  
  that bosonic and fermionic contributions to renormalization 
  of the potential terms may
  cancel each other  (as they do in the reduced model for $AdS_2 \times S^2$ 
  superstring which is the  ${\cal N}=2$ supersymmetric sine-Gordon theory). 
  Our aim below will be to present evidence that this model is indeed  UV
  finite.
  

\renewcommand{\theequation}{3.\arabic{equation}}
 \setcounter{equation}{0}

\section{Bosonic part of the reduced theory and  UV divergences}


To get an idea about the structure of  possible UV divergences 
in reduced theory \rf{lag}  let us first consider its  bosonic part.
We shall   first review  the form of the sigma model that appears 
as a result of choosing a specific  parametrization of the basic field  $g\in G $ 
 and integrating out the $H$  gauge field $A_a$.
 That assumes that the $H$-gauge is fixed by choosing a particular
form of the group element $g$. 
  
 In the case of the string on $R_t \times S^n$  or sigma model on the sphere $F/G=S^{n}$ 
 the reduced theory is  based on the 
 gWZW  model   for   $G/H=SO(n)/SO(n-1)$. It  is  constructed by  choosing a    
 parametrization of $g$  in terms of the coordinates  of the  $G/H$ coset 
  and integrating out the $H$ gauge field $A_a$. 
 We end up with an integrable theory  represented  by an  
  ($n$--1)- dimensional  sigma model with a potential (see \ci{gt}) 
 \be \la{si}
 L =  G_{mk} (x)\  \del_+ x^m \del_- x^k   -    U(x)  \ .
  \ee
 Here $x^m$  represent  the $n-1$ ($= \dim G - \dim H$) 
   independent components  of $g$  left   after fixing 
    the $H$ gauge.\foot{In contrast to the metric 
  of the usual geometric (or ``right'') coset $SO(n)/SO(n-1)=S^{n-1}$
 the metric  $G_{mk}$ in \rf{si}  found from the symmetrically gauged
 $G/H={SO(n)/ SO(n-1)}$\ gWZW  model
 will generically have singularities and no non-abelian
  isometries. The corresponding space may be denoted as  $\CS^{n-1}$.
While the gauge
  $A_a=0$ preserves the explicit  $SO(n-1)$ invariance of the equations of motion,
   fixing the gauge on $g$  and integrating out $A_a$
    breaks all non-abelian symmetries (the corresponding symmetries are
    then ``hidden'', cf. \ci{brtw}).
Instead of $R_{mk} =\ a \  G_{mk}$ for a standard sphere
 the 
 metric $G_{mk}$ satisfies $R_{mk} + 2 \nabla_m \nabla_k \Phi=0$  where $\Phi$ is
  the corresponding dilaton resulting from integrating out $A_a$.} 
The potential term  
(or ``tachyon coupling'' in string sigma model  language) 
  in \rf{si}   originates directly from
the $\km^2$ term in the action. It   is
a relevant (in the case of a compact group $F$ such as for the sphere) or
 irrelevant (in the case of a non-compact group $F$ such as for
$AdS_n$)
 perturbation of the gWZW model  and thus also of the  ``reduced'' geometry,
i.e.   it should satisfy 
  \be \la{lap}
  {1 \ov  \sqrt G e^{-2 \P} }
\del_m ( \sqrt G e^{-2 \P}  G^{mk}\del_k) U - M^2 U =0
  \ ,    \ee
  where $\P$ is the dilaton resulting from integrating out $A_a$.
An  explicit parametrization of $g$  in the case of $G=SO(n)$ 
 in terms of Euler angles is found by
 choosing  
\be \la{gah}
g=  g_{n-1} (\te_{n-1}) ... g_2(\te_2) g_{1} ( 2\vp)   g_2(\te_2) ...  
    g_{n-1} (\te_{n-1})
     \ , \ee
     where $g_m(\te) = e^{\te R_m}$  and  $R_m\equiv R_{m,m+1}$ are generators of
 $SO(n+1)$. 
Thus   $\vp\equiv \ha  \te_1,$ and $ \te_p$  ($p=2,...,{n-1}$) are
$n-1$  coordinates on the resulting  coset space  $\CS^{n-1}$, with $\vp$  
playing a distinguished  role. 
Then the  
potential $U$   has a universal  form
for { any} dimension $n$: it  is  simply proportional   to $\cos 2\vp$
  as in the sine-Gordon model ($n=2$)  \ci{gt}.
   The  metric and the dilaton
 resulting
from  integrating out the $H$ gauge 
field $A_a$  satisfy
 \be \la{dil}
ds^2 = G_{mk} dx^m dx^k=  d\vp^2 + {\rm g}_{pq}(\vp,\te)   d \te^p d \te^q \ , \ \ \ \ \ \ \ \ \ \ \ \
\sqrt G \ e^{-2 \P}    = (\sin 2\vp )^{n-2} \ , \ee
so that the equation  \rf{lap}  is indeed
 solved  by
\be \la{uu}
U= - { \mu^2\ov 2} \cos 2\vp \ , \ \ \ \ \  \ \ \ \ \ \ \ \  M^2 = - 4(n-1)  \ ,
\ee
i.e.
\be L = \del_+ \vp \del_- \vp +  {\rm g}_{pq}  (\vp, \theta) \del_+  \theta^p  \del_-\theta^q 
              +  { \mu^2\ov 2} \cos 2\vp  \ . \la{sph}
	       \ee
The explicit form of the $\CS^{n-1}$ metric \rf{dil} with   $n=2,3,4$  
 as found directly from the 
\rf{lag}  with \rf{gah} is the following. 
For the reduced models for $S^2$  and $S^3$,    i.e. for $G/H=SO(2)$ 
and  $G/H=SO(3)/SO(2)$ we have 
\be
ds^2_{n=2}=  d\vp^2 \ , \ \ \ \ \ \ \ \ \ \ \ \
ds^2_{n=3} =  d\vp^2 + \cot^2 \vp \ d\te^2 \ . \la{qwq} \ee
For $G/H=SO(4)/SO(3)$  \ci{fl} 
\be 
ds^2_{n=4}
 =  d\vp^2 + \cot^2 \vp \ (d\te_1 +  \,  V
 d \te_2)^2 +
 \tan^2 \vp\  {d \te_2^2  \ov \sin^2 \te_1 }  \ , \ \ \ \ \ \ 
 V = \cot{\te_1} \tan{\theta_2}  \ ,   \la{fol}  \ee
 or after a  change of variables  $x= \cos \te_1 \ \cos \te_2,\ \ y= \sin \te_2$
 \be 
ds^2_{n=4}  =  d\vp^2 + {\cot^2 \vp \ dx^2 +
 \tan^2 \vp  \ dy^2   \ov 1- x^2 - y^2  }  \ .   \ee
From  $G/H=SO(5)/SO(4)$  gWZW we get \ci{bs} 
\bea 
ds^2_{n=5}
 &=&  d\vp^2 + \cot^2 \vp \ (d\te_1 +  V d \te_2 + W d\te_3)^2 + 
  \tan^2 \vp\ \big( {d\te_2^2\ov \cos^2 \te_1} + {d\te_3^2\ov \sin^2 \te_1} \big)\ , 
 \la{kol}\\
 V&=& {\tan \te_1 \sin 2 \te_2 \ov \cos 2\te_2 + \cos 2\te_3},\ \ \ \ \ \ 
 \ \ \ W={ \cot \te_1 \sin 2 \te_3  \ov \cos 2\te_2 + 
\cos 2\te_3}\ . \la{pya}  \eea
Together   with the  $\cos 2 \vp$ potential \rf{uu} the latter  metric thus  defines
the reduced model  for the string on $R_t \times S^{5} $.

 One can  similarly   find the 
reduced Lagrangians   for  $F/G= AdS_n = SO(2,n-1)/SO(1,n-1)$
coset sigma models which are related to the above ones 
by  an analytic continuation.
A
``mnemonic rule'' 
to get   the $AdS_n$  counterparts of $S^n$ reduced  Lagrangians is
to change $\varphi \to i \phi$  and to reverse the overall  sign of the
Lagrangian.  In general,  that will give the 
 $G/H=SO(1,n-1)/SO(n-1)$ counterpart of \rf{sph} of the form 
\be L = \del_+ \ph \del_- \ph + \tilde {\rm g}_{pq}  (\ph, \vt) \del_+  \vt^p  \del_-\vt^q 
              -  { \mu^2\ov 2} \cosh 2\ph  \ , \la{ads}
	       \ee
where  $\tilde {\rm g}_{pq} (\ph) = - {\rm g}_{pq}  (i \ph)$
(i.e.  $\cot^2 \vp  \to    \coth^2\ph  $ in \rf{qwq},  etc.).

  The reduced model 
for bosonic strings on  $AdS_n  \times S^n$  can  then  be 
obtained  by formally  combining 
the reduced models  for strings on  $AdS_n \times S^1$ and on $R \times S^n$  \ci{gt}.
For example, in the case of a string in $AdS_2  \times S^2$  we  find 
the sum of the sine-Gordon and sinh-Gordon  Lagrangians  
\be \la{sinh}
 L = \dpp \varphi \dmm \varphi  
+ \dpp \phi \dmm \phi
+ \frac{\km^2}{2} (\cos 2\varphi  - \cosh 2\phi) \ , 
\ee
while for a string in  $AdS_3  \times S^3$   we   get (see \ci{gt,gt2})
\begin{equation}
 \label{hgi}
    L =
   \dpp \varphi \dmm \varphi  +   \cot^2 { \varphi }\ \dpp \theta \dmm \theta
   +
   \dpp \phi \dmm \phi  +   \coth^2 { \phi }\ \dpp \vt  \dmm \vt
   + \frac{\km^2}{2} (\cos 2\varphi  - \cosh 2\phi) \ . 
\end{equation}
Similar bosonic actions   are   found for a string in  $AdS_4  \times S^4$  
and in $AdS_5  \times S^5$ using \rf{fol}  and \rf{kol}.

\

Next, let us  discuss the quantum properties of the above  bosonic sigma 
models. Since
these are deformations of conformal gWZW models, we should not expect 
infinite renormalization of the resulting sigma model
metrics,\footnote{On dimensional grounds,  the deformation terms cannot
contribute to the renormalization of the two-derivative terms.}
 but the
potential terms may get renormalized. While the $\cos 2\varphi$
potential is a relevant perturbation of the coset CFT in the compact
$S^n$ case, the $\cosh 2\phi$ is an irrelevant perturbation of the
corresponding coset CFT in the $AdS_n$ case (i.e. the sign of the mass term
$M^2$ in \rf{uu} is opposite). Thus the coefficients of the two terms
in the potential in \rf{hgi} (and in similar higher-dimensional models)  ``run''
in the opposite directions. As a result, the bosonic reduced theory
like \rf{sinh} or \rf{hgi} is not renormalizable already 
at the leading one-loop order: one would need to
introduce two different bare coefficients in front of the $\cos 2\varphi $
and the $ \cosh 2\phi$ terms in the potential to cancel the divergences. 

A simple way to see  that different renormalization
is to note that  the one-loop correction given by 
$\log \det \Delta $ terms  is not sensitive to a change of 
sign of the classical action  which should be done while  going from $S^n$ to 
$AdS_n$ reduced model via $\vp \to i \phi$. Thus  
if in the $S^n$  model we get a divergence     $c_1 \cos 2 \vp \   \ln \Lambda$, then 
in the $AdS_n$  model  it  should  be given 
simply by the same  with  $\vp \to i \phi$, 
i.e.  by  $ c_1 \cosh 2 \ph \  \ln \Lambda$.
Hence  the total divergence will be  
$c_1 ( \cos 2 \vp   + \cosh 2 \ph  )\log \L $.  It will thus 
 have a  different structure than
the  classical potential in \rf{sinh},\rf{hgi}, 
and  so cannot be absorbed into  renormalization of the
single parameter $\mu$.

More generally, the supertrace  symbol  
  in  $ \mathrm{Str}(g^{-1}Tg T )$  in \rf{lag} means that
  the  potential terms  for the $AdS_5$  and 
$S^5$  parts of the  reduced theory 
are taken with the opposite signs  (i.e. as   $\cos 2 \vp - \cosh 2 \phi$ in 
the Euler  angle parametrization \rf{gah}). Since the 
 anomalous dimensions\foot{It is useful to recall that
 $\tr (g^{-1}Tg T ) $  is a  primary field of the WZW theory \ci{kz}.} 
  of the corresponding two  terms 
are opposite
(which is related to the opposite signs of curvature of 
$AdS_5$  and $S^5$),  the logarithmically divergent term  coming from 
the bosonic part of \rf{lag} 
is actually the sum, not the difference, i.e. defined   in terms of $g$ in 
the product of the two groups 
it  contains  $\mathrm{tr}$
instead of  $ \mathrm{Str}$ 
\be \la{suu}
L_{\rm 1-loop} = a_1 \  \mathrm{tr}(g^{-1}Tg T ) \ \ln \Lambda  \ . 
\ee 
One expects   that  in the full reduced theory \rf{lag}
corresponding to the 
\adss  superstring   the fermionic terms 
 will make the whole theory UV finite, i.e.
\rf{suu} will be canceled by the fermionic contributions, 
i.e. the potential  can be considered as 
an exactly  marginal perturbation  (with the value of its coefficient 
$\mu$  being finite and  arbitrary). 

This is indeed what happens  in the $AdS_2 \times S^2$ case 
where the reduced theory  is equivalent to the (2,2) supersymmetric sine-Gordon theory
\ci{gt}. 
For this to happen in the general theory \rf{lag} 
the  contribution to the  divergences coming from 
 the fermionic Yukawa interaction term 
 should also  be proportional to  \rf{suu}, i.e. to the {sum}
  of the
bosonic potentials
instead of
 their  difference entering 
the classical action.

It is possible to argue that indeed the fermionic part is invariant
under the analytic continuation $\vp \to i \ph$, so that its one-loop 
contribution to the renormalization of the bosonic 
potential should also be even, i.e.  proportional to the sum of the
potential terms as in \rf{suu}.  For example, the explicit form of the
fermionic terms in the $AdS_3 \times S^3$ case given in \ci{gt2} is
invariant under $\vp \to i \phi , \ \phi \to -i \vp$. In the next section 
we shall 
give a general argument of why 
that should happen and check explicitly
that the resulting divergent coefficient indeed cancels against the
bosonic one.

\



Let us continue with several
%
%
general remarks    about the structure  of 2-loop renormalization 
of the potential (or ``tachyon coupling'') term in a generic bosonic sigma model
\be 
S= { 1 \ov 4 \pi \a' } \int d^2 \s \ \big[  G_{mn}(x) \del^\m x^m \del_\m x^n 
+ \eps^{\m\n}   B_{mn}(x) \del_\m x^m \del_\n x^n   - U(x)  \big] 
  \ . \la{sip} \ee
The  renormalization of $\U$   is governed by the $\beta$-function
(see, e.g., \ci{cal,ts,osb,tsf})
\bea \la{beel}
&&\beta^\U =  - \g \U - 2 \U  \ , \\
&&\g = \Omega^{mn} D_m D_n +   O(\a'^4)  \ , \\ 
&&
\Omega^{mn}= \ha \a' G^{mn}  + p_1  \a'^2 R^{mn}   
+    p_2 \a'^2 H^m_{\ kl} H^{n kl}  +  O(\a'^3) \ . \la{wee}
\eea
Here we follow the notation of \ci{ts,mt2}.
The 2-loop   coefficients $p_1,p_2$   are scheme dependent (they can be changed
 by redefining
 $G_{mn}$). In dimensional regularization with minimal subtraction \ci{cal,ts}
 $ p_1=0 $ while $p_2$, in principle, 
   still  depends on how one treats $\eps^{\m\n}$ in dimensional regularization 
 (cf. \ci{mt2,bos,osb,ni}).
 In  a scheme where $\eps^{\m\n}$ is considered as  being  2-dimensional one  \ci{bos} 
 (which also   corresponds to the $f_1=-1$ scheme in \ci{mt})
  one finds \ci{mt2,osb}  $p_2 = - { 1 \ov 8} $.
  In this  case the dilaton and tachyon  2-loop $\beta$-functions
  take the form\foot{The corresponding  
 operator $\g$  enters also the dilaton  $\beta$-function  considered in \ci{mt2}.
 See also  the  discussion around eq.(5.10) in the second reference in 
  \ci{tsw}.}
  \bea \la{tww}
&&  \beta^\phi =- \g \phi +  
   { 1 \ov 6} \big[ D - { 1 \ov 4}\a'  H_{m kl} H^{m kl}   +    O(\a'^3)\big]  , \\ 
&&  \beta^\U =  - \g \U - 2 \U \ , \ \ \ \ \ \   
\g= \ha \a' \big[ G^{mn}  - { 1 \ov 4}  \a' H^m_{\ kl} H^{n kl}  +    O(\a'^3)\big] D_m D_n  \ . \la{beu}
\ee
In the case of a  WZW   model (i.e. when the group space is  a target space and $ H_{m kl}$
is the parallelizing torsion) 
these expressions  are then in agreement  with the WZW 
central charge ($ C= 6 \beta^\phi, \ \phi=\const$)
and the 
anomalous  dimension of the field $\tr g(\s)$ 
as found in \ci{kz} (see also \ci{bos,jj}): 
\be \la{cec}
C= {  k d\ov k + \ha c_{_G}}= d (1 -  { c_{_G} \ov 2 k} + ...) \ , \ \ \ \ \ \  
\g \U = { c_{_r} \ov k + \ha c_{_G}}\U = {c_{_r}  \ov k} (1 -  { c_{_G} \ov 2 k} + ...)\U \ ,     \ee
where   $\a' = {1 \ov k}, \ \   R_{mn} = {1 \ov 4} 
 H_{m kl} H_n^{\ kl} 
= {R \ov d}  G_{mn}, \ 
\ c_{_G}= { 2 R  \ov d} , \ \ G^{mn} G_{mn} = d$
and $  c_{_r} $ and $ c_{_G} $  are  the values of the 
Casimir operator  in, respectively,  the fundamental and adjoint 
  representations.  
 More explicitly, if we 
 consider the  renormalization of a potential term   in   a WZW model
 \be  L=  L_{_{\rm WZW}} (g) - \U(g)   \  , \la{wes}  \ee
as we shall do in the next section, 
then, as follows from the above general results, 
 the {\it 2-loop}  renormalization of $\U$ will 
 originate  only from the vertices in the WZ term in the action
 (and will be, in general, scheme-dependent).

 Such  a 2-loop shift in  the anomalous dimension 
 is  absent in  2d supersymmetric WZW models  due to an 
 additional contribution of the fermions that are chirally coupled to $g$.
 That  can be seen  
 by first integrating the fermions out  which leads to the shift of the overall 
  coefficient $k$  of the WZW  term\foot{In WZW model  written in
   a manifestly supersymmetric form 
 the fermions are Majorana spinors coupled to  $g$ as 
 $\tr (\bar \psi  \g_5  \g^\m [\del_\m g g^{-1} ,\psi])$, and their 
 rotation $\psi_L\to  g^{-1} \psi_L g, \    \  \psi_R\to  g \psi_L g^{-1}$
   that decouples them from $g$  produces a non-trivial jacobian
 that shifts the  coefficient of the WZW term \ci{red}.}
   $k\to k'= k- \ha c_{_G}$ 
  and thus eliminates   all higher than 1-loop contributions to the anomalous dimension 
  of $\U$: the corresponding dimension in \rf{cec} is then $ { c_{_r}  \ov k' + {1 \ov 2} c_{_G}}
  =  { c_{_r} \ov k}$. 
  
  The case of the reduced theory which we shall consider below 
  is different from the case 2d supersymmetric  WZW theory with a bosonic  potential 
  in that here there is an additional fermionic  interaction term that  
  contributes to the renormalization of the bosonic potential  and 
  completely  cancels out also  the 1-loop anomalous dimension.

\

An apparent  consequence of the above general expression  for $\beta^\U$ \rf{beu}
is that 
in the  sigma models like \rf{hgi} obtained by integrating out the gauge field $A$ 
 where there is no  WZ-type $B_{mn}$ coupling 
($H_{mnk}=0$) there will be no  non-trivial 
 renormalization of  the potential at the 2-loop order. 
  There is a caveat that  since this sigma model is obtained from a conformal gWZW 
  model its classical metric  will be  conformal  only in a  special scheme \ci{tsf};
  in a standard (minimal subtraction) 
  scheme the metric will be deformed by  $\a' = {1 \ov k}$  corrections 
  starting from the  2-loop order \ci{dvv,tsep,tsw}. As a result, 
  expressed in terms of the ``tree-level''  metric, the anomalous dimension 
  will receive an effective 2-loop contribution  coming from the 1-loop term  after 
  one uses there the  1-loop corrected metric. 
   This subtlety would be absent in a  2d supersymmetric gWZW  model where, as
    recalled above,   
  the fermions produce a 
  compensating   shift of the level $k$  and thus  the expressions for the 
  central charge,  anomalous dimension  and the effective sigma model metric 
  obtained by integrating
   out the 
  $A$ gauge field remain essentially the  1-loop ones  (see \ci{tsw} and refs. therein).
  
  Though there is no apparent 2d supersymmetry in our reduced
  Lagrangian \rf{lag} one may suspect that the effect of fermions
  there may be similar to the one in the 2d supersymmetric gWZW
  case. 
  If we
  assume that the fundamental quantum variables are actually the GS
  fermionic currents $Q_1$ and $Q_2$ in \rf{al} then \rf{psa} which
  defines $\Psi_{_L}$ and $\Psi_{_R}$ is similar to a rotation that
  decouples fermions from bosons and produces the level shift $k\to
  k'= k- \ha c_{_G}$ in the 2d supersymmetric WZW model.\foot{ Indeed,
  the standard relation \ci{pw} for a fermionic determinant implies
$ 
\det ( \del_+  + {\rm Adj}_{ g^{-1} \del_+ g}
) \det ( \del_-  + {\rm Adj}_{ \td g^{-1} \del_- \td g}
) 
= {\rm exp} [   c_{_G} I_{_{\rm WZW}} (g\td g^{-1}) ] \  \det  \del_+\  \det  \del_-.$
Here we   assumed  that fermions are in adjoint representation;
otherwise $ c_{_G}$ should be replaced by the corresponding 
quadratic Casimir of the representation, 
$ T_a T_a = c_r I$. 
This expression  can be factorized  into separate  chiral determinant contributions 
using  Polyakov-Wiegmann identity, and then 
$I_{_{\rm WZW}} (g) $  (or $I_{_{\rm WZW}} (\td g^{-1}) $) 
 can be interpreted as  the effective action for a Dirac fermion  with  purely right (left) 
 coupling to the corresponding current.
 } 
The above remark does not, however,  directly apply to our case  since 
  the fermionic kinetic term  in \rf{lag}  contains the  matrix $T$  which 
  does not in general commute with
  $g$  so  after the rotation of  $\Psi_{_L}$ we will be left 
  with a non-trivial  $g^{-1} T g$ coupling in its   kinetic term. 
 
\

As was already stressed above, compared to WZW   theory   coupled
 to fermions,
  we have in addition  a fermionic counterpart of the potential term in \rf{lag} 
  that may also  contribute to the renormalization of the bosonic potential. 
  This ``Yukawa'' interaction term originated from the fermionic WZ 
  term in the   original GS action \rf{al}
   and thus its contribution (beyond the 1-loop level) may be sensitive to a
   choice of regularization, just like  the treatment  of the bosonic WZ term is.


 These issues are   related to the fundamental question: 
 how we actually define the 
 quantum version of the reduced theory, i.e.  which is the choice of  the basic 
 quantum variables,  path integral measure  and regularization?
 This question  is especially non-trivial here in view of the absence 
 of a manifest symmetry relating the bosonic   and fermionic variables. 
 It is  natural to assume that these choices should be  made 
 so that to ensure  that the 
 resulting theory is UV finite, just like   the original GS theory  should  be. 
 
 Below we shall  assume that the fundamental fermionic variables are 
  $\Psi_{_L}$ and $\Psi_{_R}$ having canonical kinetic terms and will  show  
  that all 1-loop divergent contributions  to the potential terms cancel, 
  while the 2-loop   contributions which are, in general,  scheme-dependent, 
  also vanish  in 
   a natural  regularization scheme.

\renewcommand{\theequation}{4.\arabic{equation}}
 \setcounter{equation}{0}

\section{UV finiteness   of the   reduced theory}

In this section  we shall study  the divergences  of the  reduced model \rf{lag}
for strings in  $AdS_5\times S^5$ without first integrating out the $H$ gauge field.
This allows us to utilize explicitly the conformal invariance  of the gWZW model 
so that the  only possible renormalization 
that needs to be analyzed is that of   the potential terms.

\subsection{Change of  variables in  the  reduced  action}

To study  the quantum  properties of reduced model 
 it is useful to reorganize its  action  and decouple the 
$H$ gauge field as was already 
mentioned below eq.\rf{kuk}, i.e. following 
the same pattern as in  the bosonic gauged WZW models. 
 Namely, we  can always choose  the two-dimensional gauge fields to be of the form
\be
A^{(i)}_+=u^{(i)}\partial_+u^{(i)}{}^{-1}~, ~~~~ \ \ \ \ \ 
A_-^{(i)}={\bar u}^{(i)}\partial_+{\bar u}^{(i)}{}^{-1} \ , \la{aa}
\ee
where $i=1,2$ labels the two copies of $SO(4)$ algebra  in 
the algebra of $H$ isomorphic to $ SO(4) \times SO(4)$. Then, the
coupling between $g$ and the gauge field may be eliminated by redefining
$g =\diag ( g^{(1)},g^{(2)})   \in Sp(2,2) \times Sp(4)$ as follows
\be
{\tilde g}{}^{(i)}=u^{(i)}{}^{-1}g^{(i)} {\bar u}^{(i)}  \la{gag} \ . 
\ee
This redefinition may be written more compactly  as ${\tilde g}=    u^{-1} g \bar u$   by 
introducing the ``supermatrices'' \foot{The 
supertrace of such matrices  is defined as a  difference of traces of diagonal  blocks.}
\be\la{tap}
{\tilde g}=
\begin{pmatrix}
{\tilde g}{}^{(1)} & 0 \cr
0 & {\tilde g}{}^{(2)}
\end{pmatrix} \ , 
~~~~~ u= \begin{pmatrix}
u^{(1)} & 0 \cr
0 & u^{(2)}
\end{pmatrix}\ , 
~~~~~ {\bar u}=
\begin{pmatrix}
{\bar u}^{(1)} & 0 \cr
0 &  {\bar u}^{(2)} 
\end{pmatrix} \ . 
\ee
We can also redefine  the fermionic  fields in \rf{lag} as\foot{Note that since $u,\bar u$ are from $H$
and thus commute with $T$
the rotated fermionic fields also satisfy the constraints in \rf{coon},\rf{psis}, i.e. 
they anticommute with $T$.} 
\be
\tilde \Psi_{_L}=u^{-1} {\Psi}_{_L}u\ , 
~~~~~~~~~~~
\tilde\Psi_{_R}={\bar u}^{-1}{ \Psi}_{_R}{\bar u} \ . 
\label{chiral}
\ee
Then the reduced Lagrangian \rf{lag}  becomes 
\be
L=&&L^{(G)}_{_{\rm WZW}}({\tilde g})- k' {L}^{(H)}_{_{  \rm WZW}}(u^{-1}{\bar u})
                       +\mu^2 \Str\big({\tilde g}^{-1}T{\tilde g}T\big)\cr
 &&+\ \Str\big(  {\tilde \Psi}_{_L} T\partial_+{\tilde \Psi}_{_L}
  + {\tilde \Psi}_{_R} T \partial_-{\tilde \Psi}_{_R} \big)
  +\mu \Str\big({\tilde g}^{-1}{\tilde \Psi}_{_L}{\tilde g}{\tilde \Psi}_{_R}\big) \ . 
\label{eedS}
\ee
We used that  $u\in H$ commutes with $T$. 
Here the factor  $k'$ in the second  term  indicates the shift  of the 
overall coefficient (or the level $k$, that we  formally set to 1) 
coming from the Jacobians of the above change of variables from $A_\pm$ to $u,\bar u$ 
and from the rotations of the fermions \rf{chiral} as in  the usual 2d  supersymmetric gWZW 
case \ci{sgwzw}. 
 Here   the shift is $k'=k  + (1-  \ha) c_{so(4)} $
where $c_{so(4)}$ is  the quadratic Casimir of $H^{(1)}=SO(4)$.  The 
shift by $c_{so(4)} $  is coming from the bosonic Jacobian and by 
$-\ha c_{so(4)}$  from the 
chiral fermionic Jacobians regularized in a 
vector-like fashion so that their contributions combine
into  ${L}^{(H)}_{_{\rm WZW} }(u^{-1}{\bar u})$.

This redefinition is very useful for 
the purpose of studying 
the UV properties of the theory: we can ignore the decoupled  WZW
term  for the subgroup $H$  (i.e. the term multiplied by $k'$ in \rf{edS})
 since it is conformally
invariant on its own. 
The fermions in \rf{edS} have free kinetic terms.  By
formally assuming that $T$ transforms under $G= Sp(2,2) \times Sp(4)$
in an appropriate way\foot{One may define this transformation as
follows. The fixed matrix $T$ identifies an $SO(4)\times SO(4)$
subgroup of $Sp(2,2)\times Sp(4)$. Then, $Sp(2,2)\times Sp(4)$
transformations of $T$ amount to choosing different (but equivalent)
embeddings $SO(4)\times SO(4)\subset Sp(2,2)\times Sp(4)$. At the
level of the original action, a realization of this symmetry requires
transformations of the gauge field. This is not surprising, given that
one gauges different $SO(4)\times SO(4)$ subgroups of $Sp(2,2)\times
Sp(4)$.}  we may then treat the remaining terms in the action as being
invariant under $G$.

Let us note that in general  one can not, of course, completely decouple 
${L}_{_{\rm WZW} }(u^{-1}{\bar u})$ term:  the gauge-invariant 
observables in the original theory  may depend on $u$ and $\bar u$. 
 Indeed, the  action \rf{edS} -- even 
  written in an apparently factorized form -- still exhibits the following 
   gauge invariance
\be
{\tilde g}\mapsto h{\tilde g}h^{-1}\ , 
~~~~~~~
{\tilde \Psi}_{L,R}\mapsto h{\tilde \Psi}_{L,R}h^{-1}\ , 
~~~~~~~
u\mapsto huh^{-1}\ , 
~~~~~~~
{\bar u}\mapsto h{\bar u}h^{-1} \ , 
\ee
where $h={\rm diag}(h^{(1)},h^{(2)})\in SO(4)\times SO(4)$. 
The observables of this theory must be invariant under these transformations. 
Clearly, traces 
of products of powers of ${\tilde g}$ and $T$ are invariant. However, partial 
derivatives of ${\tilde g}$
must be promoted to covariant derivatives of ${\tilde g}$.
 Thus, $u$ and ${\bar u}$ must necessarily enter the 
observables.


\subsection{Structure of divergences in quantum effective action}

We are interested in understanding the UV finiteness properties of the theory (\ref{lag})
or, equivalently, of (\ref{eedS}).
To simplify  the notation in what follows we shall 
omit tildes on $g$  and $\Psi$ in \rf{eedS}, i.e. study
the UV properties of the following theory 
\be
L=L^{(G)}_{_{\rm WZW}}({ g})
                       + \mu^2 \Str\big({ g}^{-1}T{ g}T\big)
+\ \Str\big(  { \Psi}_{_L} T\partial_+{ \Psi}_{_L}
  + { \Psi}_{_R} T \partial_-{ \Psi}_{_R} \big)
  + \mu \Str\big({ g}^{-1}{ \Psi}_{_L}{ g}{\Psi}_{_R}\big) \ , 
\label{edS}
\ee
where $g \in Sp(2,2)\times Sp(4 )$.
 

This theory is power counting renormalizable but it 
 is not clear a priori that divergences will preserve  the specific structure of the 
 potential terms. Indeed, as was discussed in the previous section, 
 the bosonic  part  of \rf{edS}  is the  sum of the two decoupled theories for 
 $g^{(1)} \in  Sp(2,2)$ and $g^{(2)} \in  Sp(4)$ with the potential terms ``running'' 
 in the opposite directions. Thus renormalizability of the bosonic theory 
a priori would  require us  to add also the coupling (see \rf{suu}) 
$\td \mu^2 \tr\big({ g}^{-1}T{ g}T\big)$  or introduce 
two independent couplings for the
two bosonic potentials.

Moreover, fermionic coupling constant in \rf{edS} need not be equal (in
the absence of explicit 2d supersymmetry) to the square of the
coupling in the bosonic potential, i.e. it may be some $\mu'$ that may
``run'' differently than $\mu$.\foot{It is easy to see on dimensional
grounds that quartic fermionic terms (which are a priori possible to
put into the bare action) are not actually induced here with UV
divergent coefficients and thus their coefficients can be set to
zero.}  Our analysis below shows that the corresponding 1-loop
renormalization group equations admit a fixed point $\mu'=\mu, \ \td
\mu=0$, i.e.  with this choice all 1-loop divergences (including the
ones depending on fermions) cancel. As for the 2-loop divergences,
their coefficients happen, in general, to be scheme dependent and
there exists a scheme where they are absent, providing strong evidence
of the finiteness of the theory \rf{edS}.

\

We will study the divergent part of the  effective action $\Gamma[{g}]$ for 
the bosonic field $g$
obtained by expanding the fields around some generic background $ g$ (solving the classical 
equations of motion)
\be\la{ppp} 
{g} \to  {g}\ e^{\ze}\ , ~~~~~~~~{g}^{-1} \to  e^{-\ze}\ {g}^{-1} \ , 
\ee
and integrating out the fluctuation field $\ze $ (taking values in the algebra of $G$)
and the fermions. 
Let us discuss the expected structure of this  effective action.
It should   be 
consistent with all the  global symmetries  which are:

$(a)$ manifest $G=Sp(2,2)\times Sp(4)$ symmetry {\it assuming} that one treats
$T$ as a field 
transforming in the bifundamental representation. As mentioned 
above, this symmetry is  
manifest at the level of the
 classical  action (\ref{edS}).

$(b)$ symmetry under formal rescaling ${g}\mapsto  a  {g}$ which simply means 
 that each term in the
classical action contains an equal number of factors of ${g}$ and of
 ${g}^{-1}$. \foot{
Since 
${g}={\rm diag}({g}^{(1)},{g}^{(2)})$ is an element 
of $Sp(2,2)\times Sp(4)$ this formal rescaling takes us outside the domain of
definition of  ${g}$ so we will understand   this rescaling only in the sense 
of counting the numbers of  ${g}$ and  ${g}^{-1}$ factors.}

$(c)$ invariance under  ${g}\leftrightarrow {g}^{-1}$, 
${ \Psi}_{_L}\leftrightarrow{ \Psi}_{_R}$ combined with 
the world-sheet    $\sigma_+\leftrightarrow\sigma_-$ transformation.

$(d)$ ${g}^{(i)}\mapsto (-1)^{a_i} {g}^{(i)}$, 
${ \Psi}_{_{L,R}}\mapsto (-1)^{b_{_{L,R}}}{ \Psi}_{_{L,R}}$, 
with $a_1, a_2, b_{_L}, b_{_R}=0, 1$  and  $a_1+a_2+b_{_L}+b_{_R}=2$.

$(e)$ ${g}^{(1)} \leftrightarrow {g}^{(2)}  $, \ 
${\Psi}_{_L}\leftrightarrow {\Psi}_{_R}$ 
(interchanging the off-diagonal blocks in the fermionic matrices in 
\rf{exl},\rf{exz})
together with 
changing the sign of the Lagrangian, 
 i.e. the sign  of the overall  coupling  constant.

The contributions to the effective action depend on  either  $j_\pm =
g^{-1}\del_\pm g$ if they come from the WZW action or explicitly
 $g$ if they come from the $\m$-dependent (or ``deformation'')
terms in \rf{edS}. Two-dimensional Lorentz invariance requires that
all factors of the vector $j_\pm$ appear  in pairs. The structure of the
action \rf{edS}  (in particular, the chiral symmetry of the WZW model) and the
fact that $j$ has dimension 1 imply that the coefficient of the $j^2$
term must be finite (generated by diagrams containing at least two
propagators). For that reason below we will concentrate on the
derivative-independent terms built out of $g$.

The symmetries  $(a)$ and $(b)$ above imply that at each loop order the effective action 
$\G[g]$ 
is a combination of the $\tr$ and $\Str$ of polynomials in ${g}^{-n} T {g}^n T$. 
The symmetry  (c) implies that in each monomial  ${g}$ always appears raised to  the same 
power as  its inverse. The symmetry (d)  implies that the number of factors of ${g}$ 
plus the number of factors ${g}^{-1}$  in each term is even. 
Finally, the  symmetry  $(e)$ together with 
the fact that ${g}={\rm diag}({g}^{(1)},{g}^{(2)})$ is
 block-diagonal imply that 
the contribution 
to the effective action from diagrams with an {\it even}
 number of loops is the  {\it supertrace } 
of a polynomial in ${g}^{-n} T {g}^n T$ while the contribution 
 from diagrams with an {\it odd}  number of loops is the  {\it trace}  of a polynomial 
in ${g}^{-n} T {g}^n T$  (cf.  \rf{suu}).

Since the only bare ${g}$ factors may come from the potential
terms, having more than two factors of ${g}$ and ${
g}^{-1}$ requires having more than two vertices from the $\m$-dependent
terms. The number of factors of $\mu$ produced this way equals the
total number of factors of ${g}$ plus the number of factors of
${g}^{-1}$. Then the only way to obtain the correct dimension
of the effective action is to ensure that the coefficients of such
terms are given by (two-dimensional) momentum integrals with negative
mass dimension; such integrals are finite in the UV.

From the arguments above it follows that the only potentially
divergent contributions to the bosonic part of the effective action
must be proportional to $\mu^2$ {\it before} the momentum integrals
are evaluated.  Divergences of this type may be proportional to either
the bosonic potential term in \rf{lag}, i.e. $\Str[{g}^{-1} T{ g} T]$ in \rf{edS}, 
or to $\tr[{g}^{-1} T {g}
T]$. Such contributions may come from the two types of diagrams:
diagrams with one vertex from the bosonic potential and diagrams with
two vertices from the boson-fermion (``Yukawa'') interaction term in
\rf{edS}.\foot{An equivalent argument can be given of course by starting
directly with the action \rf{lag}.  Depending on the number of loops
one may have additional vertices arising from the expansion of the
action. Due to its gauge invariance, the gauge field in the gauged WZW
action can only  contribute through its field strength, so on
dimensional grounds it cannot contribute to the UV-divergent terms
proportional to $\mu^2$.}

In the following all integrals will be defined with an implicit IR 
regulator which is different from the UV regulator. This is needed
since we are interested only in UV divergences. In this regime, masses
of particles are irrelevant. In other words, we can expand in powers of 
the mass parameter of the world sheet fields or in powers  of $\mu$.

\

A special trick that we shall use below to simplify the calculation of the 
UV divergences is to treat the field  $ g$   (and the fluctuation field $\zeta$) 
as unconstrained
matrices rather than elements (of the algebra) of  $Sp(2,2)\times Sp(4)$. 
This is possible to do by assuming that the matrix multiplication in the
action  contains factors of the  symplectic $Sp(2,2)$ and
$Sp(4)$ metrics.  Such factors  project out  the non-$Sp(2,2)\times Sp(4)$ parts of the fields
in each term of the
action. Effectively,
the contraction with the symplectic metric introduces the appropriate
projectors in vertices and propagators.

To define the perturbation theory we will need the propagators for the
bosonic fluctuation fields $\ze$ in \rf{ppp} and the  fermionic fields that can be parametrized
 as ($\chi_{_{L,R}}$  and  $ \lambda_{_{L,R}}$ are $4 \times 4$ matrices  expressed in terms of 
 $\xi_{_{L,R}}$ and  $\eta_{_{L,R}}$, see \rf{ip},\rf{exl},\rf{exz}) 
\be   [T,\Psi_{_{L,R}}]=
\begin{pmatrix}
0&\lambda_{_{L,R}}\cr
\chi_{_{L,R}} & 0
\end{pmatrix}  \ .   \la{ghj} \ee 
We will use $(a,b,\dots)$  for the $Sp(2,2)$ indices and  $({\bar
a},{\bar b},\dots)$ for the $Sp(4)$ indices and introduce the  corresponding symplectic metrics  
\be
\Omega_{ac}\Omega^{bc}=\delta_a^b \ , \ \ \ \ \ \ \ \ \ \
 \Omega_{\bar a\bar c}\Omega^{\bar b\bar c}=\delta_{\bar a}^{\bar b}  \ . 
\ee
Then the bosonic propagator is 
\be
\langle \ze_{ab}\ze_{cd}\rangle =\frac{a_{_b}}{p^2}
(\Omega_{ac}\Omega_{bd}+\Omega_{ad}\Omega_{bc}) \ , \  \ \ \ \ \ \ \ \ \ 
\langle \ze_{{\bar a}{\bar b}}\ze_{{\bar c}{\bar d}}\rangle =- \frac{a_{_b}}{p^2}
(\Omega_{{\bar a}{\bar c}}\Omega_{{\bar b}{\bar d}}+\Omega_{{\bar a}{\bar d}}\Omega_{{\bar b}{\bar c}})
\label{bo} \ , 
\ee
and the fermionic one is ($p_\pm=p_0\pm p_1$)
\be
\langle \lambda_{_L}{}_{a{\bar b}}\chi_{_L}{}_{{\bar c}d}\rangle=
\frac{i\,a_{_f}}{p_{+}}
(T_{ad}\Omega_{{\bar b}{\bar c}}-T_{{\bar b}{\bar c}}\Omega_{ad}) \ , 
\  \ \ \ \ \ \ \ \ \   
\langle \chi_{_R}{}_{{\bar c}d}\lambda_{_R}{}_{a{\bar b}}\rangle=
\frac{i\,a_{_f}}{p_{-}}
(T_{{\bar c}{\bar b}}\Omega_{da}-T_{da}\Omega_{{\bar c}{\bar
b}})~  \ . \la{fe} 
\ee
Here $a_{_b}$ and $a_{_f}$ are normalization constants
\be \la{norma}
a_{_b}=- {1 \ov 4}\ ,\ \ \ \ \ \ \ \ \ \    \ a_{_f}= {1 \ov 2}        \ , \ee 
which  we shall sometimes  keep arbitrary for generality.

\subsection{1-loop  order }

The 1-loop contribution to the effective action $\G[g]$  
 is given simply by the logarithm of the ratio of
 the determinants of 
the bosonic and fermionic kinetic operators in the $g$-background.  
To test  its finiteness it is enough to show 
the cancellation of the first two terms in the $\mu$-expansion of the logarithm of
these determinants. 

\begin{figure}[ht]
\centerline{\includegraphics[scale=0.5]{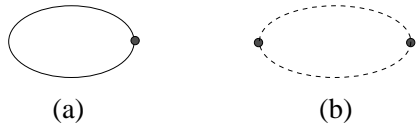}}
\caption{One-loop diagrams  contributing to the logarithmic divergences. 
 Bosonic propagators are denoted by solid lines
 and fermionic ones  by dashed lines. Black dots denote vertices coming from the bosonic and 
 the bosonic-fermionic potential term in the classical action \rf{edS}.} 
 \label{fig_1loop}
\end{figure}

The leading ($\mu$-independent  power-like divergent) term in the expansion 
 simply counts the difference between the number of 
bosonic and fermionic degrees of freedom and    thus 
 cancels automatically. 
To demonstrate  the  cancellation of the subleading (logarithmic) divergence requires a short calculation. 
The relevant Feynman diagrams are shown  in figure \ref{fig_1loop}.

 These diagrams represent the next-to-leading order in the mass  $\mu$ 
  expansion of the trace  of logarithm of the bosonic and fermionic kinetic operators.
   Their cancellation tests the mass sum rule for the  fluctuation  fields 
\be
\sum_i(-1)^{f_i}m_i^2=0\ .
\label{sumsq}
\ee
The vertices  in figure \ref{fig_1loop} arise from the expansion of 
the bosonic and the  fermionic terms in the action \rf{edS} 
 ($g$ here is the background field)
\bea
&&{ L}^{(b)}_{2}=\ha \mu^2\Str\left[  \left( \ze^2 T + T \ze^2-2 \ze T\ze\right)g^{-1}Tg\right] \ , 
\la{kp} \\ 
&&{ L}^{(f)}_{2}=\mu\Str\left[\Psi_{_R}g^{-1}\Psi_{_L}g\right] \ . \la{ghl}
\ee
We shall formally assume  that the  fields have 
$Sp(n-2, 2)\times Sp(n)$-valued indices (we will set $n=4$ at the end). Then  
the relevant contribution of the bosonic  diagram to the effective action is
\be
L^{(b)}_{\rm 1-loop}=\mu^2 a_{_b}
\big({\textstyle {n+1\ov 2}}+{\textstyle {n+1\ov 2} }-1\big)\  I_1 \ 
\tr[g^{-1}TgT]\ , \ \ \ \ \  \ \ \ \ \ \ 
I_1 = \int \frac{d^2p}{(2\pi)^2}\frac{1}{p^2} \ , 
\label{bose1}
\ee
where $\tr$ is the {\it trace} over $Sp(n-2, 2)\times Sp(n)$ indices  and 
in the integral $I_1$ we assume  the presence of both UV and IR 
cutoffs.\foot{Thus the 1-loop 
bosonic anomalous  dimension of the operator  $\tr[g^{-1}TgT]$ in $G= Sp(n)$ WZW theory 
is 
proportional to $n$.  This coefficient is  different from the dimension of 
$\tr g$    which is proportional to $n+1$  (see \rf{cec}, $c_{_r} (Sp(n)) = n+1$).
From the general perspective of the sigma model anomalous dimension 
in \rf{beu}   this   difference can be attributed to the difference  of eigenvalues of 
the  Laplace operator on the group space when acting on   the corresponding operators.
To compute the  action of the Laplacian   on $\tr[g^{-1}TgT]$ one  may follow 
\ci{jj} and use that $\del_a  g = g E^m_a T_a$  and $T_a T_a = c_{_r}  \id$
as well as a relation for $ T_a [T ,T_a]$   similar to the one appearing  in 
\rf{exp_vals} (this additional contribution  leads to the subtraction of 1 from $c_{_r}  = n+1$).}
In what follows we shall use dimensional ($d=2-2\vep$)  UV regularization, and the 
IR divergences   can be subtracted as, e.g., in \ci{gkz,ni}  by 
replacing the massless propagators by ${1\ov p^2} \to {1\ov p^2} + 
{\pi \ov \vep} \delta^{(2)}(p)$.

The three terms in the bracket in \rf{bose1}
came from the three terms in ${ L}^{(b)}_{2}$ in \rf{kp}. We used that (cf. \rf{bo}) 
\be
&&
\langle\ze^2_{ad}\rangle\equiv
\langle\ze_{ab}\Omega^{bc}\ze_{cd}\rangle=\frac{a_{_b}}{p^2}(1+n)\Omega_{ad}\ , 
~~~~~~~
\langle\ze^2_{{\bar a}{\bar d}}\rangle\equiv 
\langle\ze_{{\bar a}{\bar b}}\Omega^{{\bar b}{\bar c}}\ze_{{\bar c}{\bar d}}\rangle
=-\frac{a_{_b}}{p^2}(1+n)\Omega_{{\bar a}{\bar d}}\ , 
\cr
&& \langle (\ze T \ze)_{ah}\rangle \equiv \langle \ze_{ab} \Omega^{bc} T_{cd} \Omega^{de} \ze_{eh} 
\rangle =-\frac{a_{_b}}{p^2}T_{ah} \ , 
\label{exp_vals}
\ee
%
%
and  that $T$ with {\it two lower indices} (i.e. with one index lowered by  $\Omega$)
 is an antisymmetric matrix. 
The fermionic contribution is 
\be
L^{(f)}_{\rm 1-loop}=\ha \mu^2 a_{_f}^2 \big(n+n \big)\ I_1\  \tr[g^{-1}TgT]\ , 
\la{fea}
\ee
where in the denominator of the integral  we used that  $- p_+ p_-= p^2$  and 
the overall $\ha $ came form the expansion of the logarithm of the kinetic operator 
to the 
 second order.  To arrive at \rf{fea} we noted that 
decomposing each vertex in $4\times 4$ blocks transforming in the representations 
of $Sp(n-2,2)\times Sp(n)$ 
one finds two terms for each vertex. Each term in one vertex contracts with 
exactly one term in the 
second vertex and each contraction yields one of the two terms
 in the bracket in \rf{fea}.

Adding $L^{(b)}_{\rm 1-loop}$ \rf{bose1} and $L^{(f)}_{\rm 1-loop}$ \rf{fea}
one observes that they cancel out
(since  according to \rf{norma} \   $a_{_b}=-a_{_f}^2=-{1 \ov 4}$). This 
implies  that
(\ref{sumsq}) is indeed satisfied and thus the  1-loop effective action  for 
a generic classical  background $g$ is finite.

\

Similarly, one may show also the non-renormalization of
the fermionic interaction term in \rf{edS}, implying the cancellation
of the 1-loop correction to the fermionic propagator.  Note that in
the $AdS_2 \times S^2$ case the presence of two-dimensional
supersymmetry in the reduced action \ci{gt} makes this calculation
redundant, but in general we do not know which symmetry (if
any) relates the bosonic and the fermionic potential
terms in the reduced Lagrangian \rf{lag}. Since these two terms
appeared  (after gauge fixing and field redefinition) from the original GS
action \rf{al} where their coefficients were related by $\kappa$-symmetry this
non-renormalization effectively checks the consistency of the
reduction procedure at the quantum level.

To  check   that there is no renormalization of the fermionic potential in \rf{edS}
we should consider the   diagram containing  a single
bosonic loop and an interaction vertex coming from the expansion of
the fermionic interaction term   to second order in the  bosonic fluctuations:
\be
L^{(f)}_{\rm int} = \mu \Str\Big[g^{-1}\Psi_{_L}g\;\left(
\ze^2\Psi_{_R}-\ze\Psi_{_R}\ze+\Psi_{_R}\ze^2\right)\Big] \ . \la{kq}
\ee
The bosonic propagators (\ref{bo}) and the fact that the fermions
transform in the bifundamental representation of $Sp(2,2)\times Sp(4)$
imply that the expectation value of the second term in the bracket in \rf{kq}
vanishes
identically. Finally, the sign difference between the expectation values
in the first line of equation (\ref{exp_vals}) implies that the
contributions of the remaining two terms cancel each other. Indeed, we get 
\be
&&(g^{-1}\Psi_{_L}g)^{a{\bar b}}\left(
  \langle(\ze^2)_{{\bar b}{\bar c}}\rangle\Psi_{_R}^{{\bar c}d}\Omega_{da}
+\Omega_{{\bar b}{\bar c}} \Psi_{_R}^{{\bar c}d}\langle(\ze^2)_{da}\rangle\right)\cr
&&- \ 
(g^{-1}\Psi_{_L}g)^{{\bar a}b}\left(
  \langle(\ze^2)_{{b}{c}}\rangle\Psi_{_R}^{c{\bar d}}\Omega_{{\bar d}{\bar a}}
+\Omega_{{b}{c}} \Psi_{_R}^{c{\bar d}}\langle(\ze^2)_{{\bar d}{\bar a}}\rangle\right)
\ee
where each line represents one of the two terms of the supertrace,  
and then the
sign difference 
between $\langle(\ze^2)_{{\bar b}{\bar c}}\rangle$   and
$\langle(\ze^2)_{da}\rangle$   in  (\ref{exp_vals})
 implies that each parenthesis vanishes identically.


\subsection{2-loop order }

Let us now 
proceed to analyzing the 2-loop divergent contributions to the action
 in \rf{edS}.
We shall ignore the power divergences.\foot{They  are  absent in dimensional regularization 
and in any
case  should cancel  due to the balance of degrees 
of freedom, the mass sum rule \rf{sumsq}  or under 
 an appropriate choice of the path integral measure.}
The $\ln^2 \Lambda$ (or double-pole)  divergences should cancel 
(according to the standard  argument)  due to the cancellation of the 
logarithmic divergences at the 1-loop order established above. 
The main issue  will thus be  the   $\ln \Lambda$ (or single  pole) 
divergences.
We shall first consider  corrections to bosonic potential
and then  discuss possible  divergent contributions to the fermionic Yukawa term.

\subsubsection{Contributions to bosonic potential \label{sec:bose_pot}}

\begin{figure}[ht]
\centerline{\includegraphics[scale=0.4]{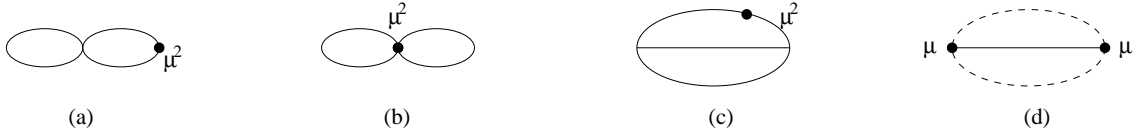}}
\caption{Two-loop diagrams at order $\mu^2$. 
Bosonic propagators are denoted by solid lines
 and fermionic ones  by dashed lines.} 
 \label{fig_2loops}
\end{figure}

The relevant 
diagrams (that may produce potentially divergent order $\mu^2$ contributions)  contain 
 one  $\mu^2$-vertex from 
the bosonic potential or  two $\mu $-vertices from the bosonic-fermionic  interaction term;
they   are shown in figure \ref{fig_2loops}.

The first diagram contains one parity-even 4-point vertex from
$L_{_{\rm WZW}}$  in \rf{edS}  (we shall suppress the  overall $k\ov 4\pi$ factor)
 \be \la{ffg}
 L_{_{\rm WZW}}{}_{_{ (4)}}=-\frac{1}{12}\eta^{\m\n}\STr\left[[\partial_\mu\zeta,\,\zeta] \ , 
[\partial_\nu\zeta,\,\zeta]\right] \ee
and an insertion of a 2-point vertex from the bosonic potential (``mass insertion''). 
As in \rf{kp}  in eq.\rf{ffg}  $\zeta$ is assumed to be a  matrix in the algebra of 
$Sp(n-2,2)\times Sp(n)$ (we will again set $n=4$  at the end).  
Namely, it is  a symmetric matrix when written with both lower indices
(i.e. with the upper  index contracted   with the 
 symplectic metric $\Omega$).
The  corresponding   contribution  to the effective action is proportional  to 
the  tadpole integrals:
\bea
&& L_{\rm 2-loop}^{(a)}=\hbar \frac{2\mu^2}{3}a_{_b}^3 \ n(n+2)\   [I_1(\vep)]^2\ \Str[g^{-1}TgT] \  , 
\label{figa}\\
&&I_1(\vep)\equiv \int \frac{d^dp}{(2\pi)^d}\frac{1}{p^2}~~, \ \ \ \ \ \ \ \ \ \ d= 2 - 2\vep \ . 
\la{ii1}\eea 
$\hbar \equiv { 4 \pi \ov k}$ is the inverse of the 
coefficient  in front of the classical WZW action. 
We  assumed   dimensional 
regularization.\foot{As we are interested in isolating
 the UV divergence, we 
 understand this integral as having an implicit IR cutoff separate 
from the dimensional regulator, e.g.,   one may carry out an IR
subtraction at the level of the propagators  as was already mentioned above.} 

The second diagram, containing one vertex from the bosonic
potential, also yields only tadpole integrals. 
The bosonic 4-vertex arising from the expansion of the  bosonic
potential is 
\be\la{fs}
{ L}_{_{\rm pot}}{}_{_{ (4)}}=\mu^2 \Str \Big(\Big[
\frac{1}{4!}(\ze^4 T + T \ze^4)-\frac{1}{3!}(\ze^3 T\ze + \ze T \ze^3)
+\frac{1}{(2!)^2}\ze^2T\ze^2 \Big]  \,g^{-1}Tg\Big) \ , 
\ee
where the multiplication of matrices is assumed with the symplectic
metric.  Also, the propagator in (\ref{bo}) enforces the condition
that $\ze$ belongs to the algebra of $Sp(n-2,2)\times Sp(n)$.  As was
already mentioned above, this implies that we may formally treat $\ze$ as
an unconstrained matrix rather than an element of the algebra of
$Sp(n-2,2)\times Sp(n)$.

The contribution of each of the three terms in \rf{fs}
to the divergent part  in the case when the group
is  $Sp(n)$  is proportional to
\be \la{bee}
2\times\frac{1}{4!}(n+1)(2n+1)-2\times \frac{1}{3!}[-(2n+1)]
+\frac{1}{(2!)^2}[(n+1)^2-(n+1)+1] ~.
\ee
This expression holds also if  we replace $Sp(n)$ by  $Sp(n-2,2)$.
Then the resulting  divergent contribution  to the  bosonic potential  term in 
the  effective action  is 
\be
L^{(b)}_{\rm2-loop}=\hbar \frac{\mu^2}{12}a_{_b}^2\ n(5n-2)\  [I_1(\vep)]^2 \
\Str[g^{-1}TgT]  \ . 
\label{figb}
\ee
As was already mentioned above, while at odd number of loops the
divergent contributions from individual
diagrams are proportional to $\mu^2 \tr[g^{-1}TgT]$, at
even number of loops the divergent contributions
  are proportional to $\mu^2 \Str[g^{-1}TgT]$, i.e. have the
same form as the classical potential.

\


Next, there is a divergent contribution from a diagram (c) with two
cubic vertices from the WZ term in the WZW Lagrangian \rf{edS} and
with a $\mu^2$ insertion from the potential. Up to a normalization
factor $\hbar^{-1}=\frac{k}{ 4\pi} $ common to the parity-even part of
the WZW Lagrangian, the cubic interaction term is
\be\la{klp}
L_{_{\rm WZW}}{}_{_{ (3)}}=\frac{2}{3}\,\epsilon^{\mu\nu}\STr[\zeta\,
\partial_\mu\zeta\,\partial_\nu \zeta]  \ .  
\ee
It then yields
\bea
&&  L^{(c)}_{\rm2-loop}=16\hbar \mu^2 
a_{_b}^4 \  n(n+2) \  I_2\  \STr[g^{-1}TgT]  \ ,\label{figc} \\
&& I_2 \equiv -
\int\frac{d^d p d^d
q}{(2\pi)^{2d}}\frac{(\epsilon^{\mu\nu}p_\mu q_\nu)^2}{p^2 q^2
[(p+q)^2]^2}\ . \la{i2}
\eea
Here we again  assumed continuation $d=2-2\vep$  but we  
need to decide how to treat $\epsilon_{\mu\nu}$ in dimensional
regularization. 
This is a well-known issue (see, e.g., \ci{fv,mt2,bos,fuch,ni}).
In general, different regularization prescriptions may lead to different 
results -- the coefficient of the 
2-loop logarithmic divergences may be scheme-dependent, with different 
results related by redefinitions of the coupling constants \ci{mt2,aj}.

Similarly  to the original GS action \rf{al} containing the
fermionic WZ term, the reduced action \rf{lag} or \rf{eedS} does not
admit a straightforward $d$-dimensional generalization.
This is analogous to (chiral) supersymmetric theories (see, e.g., 
\ci{gg,fuch,lam}) where it is 
natural to use the version of {\it dimensional regularization by
dimensional reduction} \ci{sieg}. 
We shall discuss alternative regularization schemes in 
Appendix~A and draw an analogy with the case of 2d
supersymmetric sigma models in Appendix~B.

Under this prescription we shall do all Lorentz (and spinor) algebra
in 2 dimensions and continue to $d$ dimensions only 
scalar momentum integrals. In particular, we shall use the
2-dimensional relation
\be\la{eee}
\eps^{\m\n} \eps^{\m'\n'}= - \eta^{\m'\m} \eta^{\n' \n} 
                           + \eta^{\n'\m} \eta^{\m'\n}
\ , \ee 
where in  the Minkowski signature notation
$\eta^{\m\n}=(-1,1)$.
Under this  prescription 
\be \la{eeee}
-(\epsilon^{\mu\nu}p_\mu q_\nu)^2 =
p^2 q^2 - (p\cdot q)^2 \ , 
\ee 
and thus continuing to $d=2-2\vep$  dimensions we find
\be \la{onr}
I_2  = \int\frac{d^dp d^d q}
{(2\pi)^{2d}}\frac{p^2q^2-(p\cdot q)^2 }{p^2 q^2 [(p+q)^2]^2}=
\frac{1}{4}[I_1(\vep)]^2  \ . 
\ee
The contribution of the  diagram  (c) in (\ref{figc}) is then given by 
\be 
\la{ccc}
 L^{(c)}_{\rm2-loop}=
 4\hbar \mu^2 a_{_b}^4 \  n(n+2) \ 
  [I_1(\vep)]^2       \
\STr[g^{-1}TgT]  \ . 
\ee 
Adding together (\ref{figa}),(\ref{figb}) and (\ref{figc}) and using
that $a_{_b}=-{1\ov 4}$ we find that the contribution of the bosonic
2-loop diagrams to the UV singular part of  2-loop effective
Lagrangian is
\be
L^{\rm bose}_{\rm 2-loop} = \hbar\frac{\m^2}{32}\ n^2 
\ [I_1(\vep)]^2\  \STr[g^{-1}TgT] \ , 
\label{bose2}
\ee
 where 
\be  \la{exx}
[I_1(\vep)]^2 =  \Big[ {1 \ov 4 \pi \vep} + {\cal O}(1) \Big]^2 =  {1 \ov (4 \pi)^2 \vep^2 }  + ... 
\ . \la{eqa} 
\ee
The coefficient of the most singular term
is consistent with the expected renormalization group behavior of the 
bosonic theory,
i.e. it is related to the square of the coefficient of the 1-loop single-pole
in \rf{bose1}. The coefficient of the 2-loop subleading $1\ov \vep$
pole is, in general, scheme dependent; in the standard minimal
subtraction scheme we then get no genuine 2-loop divergence (i.e. the
2-loop anomalous dimension coefficient vanishes).


Let us introduce the renormalization constant $Z^{(i)}$, $i=1,2$, for
the two bosonic operators $\U^{(i)}$ corresponding to two factorized
parts (related to the two subgroups of $Sp(n-2,2) \times Sp(n)$) in $
\m^2 \STr [g^{-1}TgT]$, i.e.  $\U^{(i)} = Z^{(i)} \U_{\rm bare}^{(i)}$
\be \la{zez}
Z = \m^{-2 \vep } \Big[  1 +  \hbar { \g_1 \ov \vep} + \hbar^2   
 \big(\frac{\gamma_2}{2\vep}  + \frac{\gamma_1^2}{2\vep^2}\big)+ ... \Big]   \ , \ee
where we suppressed the index $i$ and we have chosen $\m$ to  be 
the renormalization scale parameter.
 Then the corresponding anomalous dimension is
\be  \la{zezz}
\gamma=   { d Z^{-1}  \ov  d \ln \m } = 2 \vep +  \hbar  \g_1 + \hbar^2  \gamma_2 + ...\ .   \ee
From (\ref{bose1}) it is easy to see that $\gamma^{(1,2)}_1= \pm {1
\ov 16 \pi} n$ which, when squared,
 reproduces the coefficient of the $1\ov \vep^2$ pole in
 (\ref{bose2}).

\

Let us now consider the fermionic contributions to the 2-loop
divergent part of the bosonic effective action. There are several
types of $\mu^2$ terms which arise from bose-fermi interaction term in
\rf{edS} and they correspond to the diagrams $2(d)$ and $2(e)$. They
can be represented symbolically as coming from the square of the
interacting terms in the action:
\be
&&2\times \frac{1}{2} \langle  \int d^2 \s\
\Str[g^{-1}\Psi_{_L}g\Psi_{_R}] 
\ 
 \int d^2 \s\ \Str\Big[g^{-1}\Psi_{_L}g\;\left(
\ze^2\Psi_{_R}-\ze\Psi_{_R}\ze+\Psi_{_R}\ze^2\right)\Big]  \rangle  
\cr
&&+\ \frac{1}{2}  \langle \Big(\int d^2 \s\
\Str[g^{-1}\Psi_{_L}g\left(\ze\Psi_{_R}-\Psi_{_R}\ze\right)]\Big)^2
\rangle  
\ . 
\label{fev}
\ee 
The terms in the first line, diagram $2(d)$, lead to vanishing
contributions to the logarithmic divergences either because of
impossibility of proper Wick contractions (as in the second term in
the brackets) or because of $\Str {\bf 1} =0$ (as in the case of the
first and the third term).\footnote{This is essentially the same
calculation which implies the non-renormalization of the fermionic
potential at 1-loop order.}  The  remaining non-trivial contribution comes
from the term in the second line of \rf{fev}, i.e. diagram $2(e)$
\bea\la{ffq}
&&L^{(e)}_{\rm2-loop}= \hbar  
\mu^2 a_{_b} a_{_f}^2 \ 2\times\frac{1}{2}\left[n(n+1)-n\right]\ 
  I_3 \   \Str[g^{-1}TgT] \ , \\ \la{tri}
&&I_3= \int {d^2 p\ov (2 \pi)^2 }{  d^2 q\ov (2 \pi)^2 } 
\frac{p_+q_-}{p^2q^2(p+q)^2} \ , 
\eea
where we took into account the  minus  signs  due to the fermionic loop, 
due to the supertrace in eq.(\ref{fev}) and  due to 
the factors of $i$ in the fermionic propagators.  The $\ha $ factor is
inherited from the last line of eq.(\ref{fev}) and the overall factor
of 2 is present  because the relevant  contribution comes from the
cross term in the square.


Here again there is an ambiguity in 
defining the integral $I_3$,
i.e. 
in extending the factor $p_+ q_- $ in the integrand (which has
its origin in the chiral nature of the fermion coupling in \rf{edS})
to $d$ dimensions. In the GS action, the fermionic current components
were 2d vectors and they were reinterpreted as 2d Weyl spinors in the
reduced theory.  The fermionic interaction term in the reduced theory
\rf{lag} originated from the WZ term in the GS action \rf{al}, which
suggests that chiral fermions should be treated as if they were
2-dimensional fields. An analogy with the 2d supersymmetric gWZW model
suggests again
to  use the  regularization by dimensional reduction.

Explicitly, that means  that  we shall first use that in 2 dimensions 
 \be 
\la{ttq} 
p_+ q_- = (p_0 + p_1) (q_0-q_1) = - ( \eta^{\m\n} + \eps^{\m\n}) p_\m q_\nu 
 \ . 
\ee 
 Equivalently, interpreting $\Psi_{_L}$ and $\Psi_{_R}$ in \rf{edS} as
 upper/lower components of left/right MW 2d spinor and rewriting the
 fermionic terms using the 2-component notation with the explicit 2d
 $\gamma$-matrix factors we observe that $p_+q_-$ in $I_3$ in \rf{ffq}
 arises from
\bea 
p_+q_- = -\Tr[p\llap/ q\llap/ \ha(1+\gamma_{3})]= -  
 { p}\cdot{ q}  -  \eps^{ \mu \nu}{ p}_{\mu}{ q}_\nu  \ , 
\la{oyt}\ee
where  $\gamma_{3}= \g_0 \g_1$  and we assumed  that all spinor algebra 
is done in 2 dimensions.\foot{Same result for the parity-even term 
is found if we extended momenta and $\gamma$-matrices 
to $d$ dimensions by  assuming  that 
$
p\llap/={\bar p}^\mu{\bar \gamma}_\mu+{\hat p}^\mu{\hat \gamma}_\mu~,$  ~~
$ \{{\bar\gamma}^\mu,\,\gamma_{3}\}=0~,$  $~
[{\hat\gamma}^\mu,\,\gamma_{3}]=0 $, where  $\bar \m$ are 2-dimensional 
and $\hat \m$ are $-2\vep$ dimensional indices, i.e. $\mu=(\bar \m, \hat \m)$.}

Observing that the term with a single factor of the antisymmetric
tensor $\eps^{\m\n}$ can not contribute to the integral and continuing
the scalar integrand to $d$ dimensions we end up with
\be I_3&=& 
%
%
-\int {d^d pd^d q}\frac{p\cdot
q}{p^2q^2(p+q)^2} =-\frac{1}{2}\int {d^dpd^dq}
\frac{(p+q)^2-p^2-q^2 }{p^2q^2(p+q)^2}\cr
&=&\frac{1}{2}[I_1(\vep)]^2     \ . \la{irs}
\ee 
Then finally 
 (using \rf{norma})
\be\la{hopi}
L^{\rm fermi}_{\rm2-loop}=  - \frac{1}{32} \hbar \mu^2 \ 
n^2  \   [I_1(\vep)]^2 \ \Str[g^{-1}Tg T]  \ .
\ee
Combining this with the bosonic contribution in
\rf{bose2} we conclude that the two contributions  cancel each other, i.e. 
the  bosonic  part of the 2-loop  effective action is UV finite, 
\be
L_{\rm 2-loop}^{(\rm bos.pot.)}=
L^{\rm bose}_{\rm 2-loop}+L^{\rm fermi}_{\rm 2-loop}= {\rm finite}
\ . \la{pqw} \ee 
As already mentioned above, this is just a reflection of the
cancellation of the 1-loop logarithmic divergences as all simple $1\ov
\vep$ poles in both the bosonic and the fermionic contributions
computed in the dimensional reduction scheme come together with a
$1\ov \vep^2$ pole which  is controlled by the 1-loop divergences.
%
%
%


\subsubsection{Contributions to  fermionic potential  term }

%
The above observation, that the 2-loop correction to renormalization
of the bosonic potential is scheme dependent, may seem to contradict
the standard lore:
in view of  the cancellation of the one-loop renormalization of the 
potential, one could expect  that the two-loop renormalization
should  be scheme independent  being the first  non-vanishing correction. 
However, as discussed in section 3 and below eq.\rf{edS}, 
the reduced theory,  when viewed as a power-counting renormalizable 
model,  is actually  a multi-coupling theory 
(with the level $k$ and several $\mu$-parameters as its couplings, with the 
action \rf{edS} corresponding to a fixed-point choice). In such a case 
the 2-loop  anomalous dimension coefficients may still  be scheme-dependent.

As was already mentioned, several a priori distinct parameters in the action 
were set to be equal
as required by the reduction procedure starting from the GS action
where they were related by symmetries.
 In the bosonic part of the theory these were the couplings of the two
 potential terms corresponding to $Sp(n-2,2)$ and $Sp(n)$. With
 fermions included,
the coefficients of the bosonic  and the 
fermionic potential terms, $\STr[g^{-1}TgT]$ and
$\STr[g^{-1}\Psi_Lg\Psi_R]$, were also  related.
It is then necessary to ensure that such  relations survive 
quantum corrections. 

As we have found above, the corrections to the bosonic potential are
finite in a special dimensional reduction scheme. Finiteness of the
full theory then requires that corrections to the fermionic potential
be finite in that same scheme. In the apparent absence of worldsheet
supersymmetry which would relate the bosonic and the fermionic
potentials (and thus their renormalization, assuming one uses a
supersymmetry-preserving regularization scheme) this is not a priori
guaranteed.\footnote{It is, however, important to recall again that
the bosonic and the fermionic potentials are closely connected to the
kinetic and WZ terms in original Green-Schwarz action where the
relation between their coefficients is a consequence of the
$\kappa$-symmetry. It is possible that a global remnant of the
$\kappa$ symmetry that may be surviving in the gauge \rf{iu} offers a
sufficient protection to guarantee this relation to all orders in
perturbation theory in the reduced model.}

It is therefore crucial to test the finiteness of the corrections to
the fermionic potential in \rf{edS}
\be
U_f=\mu\STr\big(g^{-1}\Psi_{_L} g\Psi_{_R}\big)~  \la{vf}
\ee
in the  same dimensional reduction  scheme.

On dimensional grounds, to (logarithmically) renormalize $U_f$
 we need terms with a single power of $\mu$. Since all the fermionic
 interactions in \rf{edS} are proportional to $\mu$ and the bosonic
 potential is proportional to $ \mu^2$, it follows that this
 renormalization is entirely governed by the bosonic $Sp(n-2,2)\times
 Sp(n)$ WZW model with fermions treated as background fields.

\begin{figure}[ht]
\centerline{\includegraphics[scale=0.4]{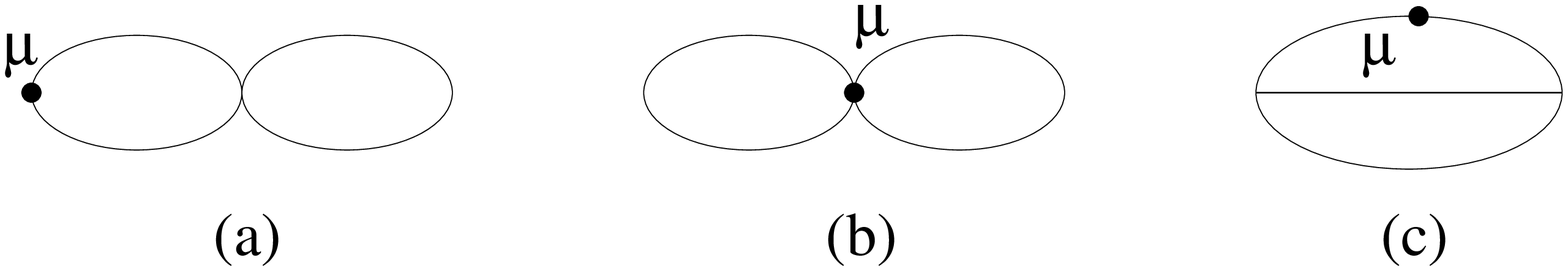}}
\caption{Two-loop diagrams 
contributing to renormalization of 
the fermionic potential. Solid   lines are bosonic propagators 
and external fermionic legs at each $\mu$-vertex are suppressed. 
} 
 \label{fig_2loops_fermi_pot}
\end{figure}

The relevant diagrams are shown in figure \ref{fig_2loops_fermi_pot}. 

The computation of their divergent parts is formally similar to that
of the renormalization of the bosonic potential in \rf{edS}, assuming
one treats $T$ as a background field. There are, however, certain
differences
related to the different  algebraic structure of $T$ and
$\Psi$, which prevent the bosonic results from being immediately used
here. 
Nevertheless, the mere fact that the calculation is effectively governed by
the undeformed $Sp(n-2,2)\times Sp(n)$ WZW model  guarantees already 
that the same scheme dependence which entered the bosonic calculation
will enter here as well.

Upon using the fact that $\Psi_{_{L,R}}$ are off-diagonal
(transforming in bi-fundamental representation of $G$, see
\rf{exl},\rf{exz},(\ref{ghj})) and that $g$ is diagonal
(cf. \rf{tap}), it is easy to see that the fermionic potential may be
written as
\be
U_f=\mu\Big(\Tr[g^{(1)}{}^{-1}\lambda_{_L}  g^{(2)}\chi_{_R}] - 
        \Tr[{g^{(2)}}{}^{-1}\chi_{_L} g^{(1)}\chi_{_R}]\Big)~~,
\ee
where $g^{(1)}\in Sp(n-2,2)$ and ${ g^{(2)}}\in Sp(n)$. 
Since the $Sp(n-2,2)$ and $Sp(n)$ WZW models are coupled only through the 
$\mu$-dependent fermionic  terms, it follows that, for the purpose of the
renormalization of $U_f$, we may treat $g^{(1)}$ and ${ g^{(2)}}$ separately. 
Thus, in a diagram of topology \ref{fig_2loops_fermi_pot}(a) the
fields propagating in the two loops must be of the same type since the
quartic vertex coming from the WZW action involves fields of only one
type (there are two distinct diagrams in this class). In a diagram of
topology \ref{fig_2loops_fermi_pot}(b) the fields propagating in the
two loops may be either of the same type or of different types (there
are three distinct diagrams in this class).  In a diagram of topology
\ref{fig_2loops_fermi_pot}(c) the fields propagating in the two loops
must be of the same type (there are two distinct diagrams in this
class).

The diagrams of these three topologies contribute as follows to the
2-loop effective Lagrangian:
\be
&&L_{\rm 2-loop}^{(a)}=\hbar  \mu\big[
\frac{1}{3}a_{_b}^3(n+1)(n+2)+(-1)\frac{1}{3}(-a_{_b})^3(n+1)(n+2)
\big] [I_1(\varepsilon)]^2
\ \STr[g^{-1}\Psi_{_L} g\Psi_{_R}]\;
\cr
\label{fermi_b}
&&L_{\rm 2-loop}^{(b)}=
-\hbar \mu\frac{a_{_b}^2}{12}(n+1)(n+2) [I_1(\varepsilon)]^2\
 \STr[g^{-1}\Psi_{_L} g\Psi_{_R}] 
\\
\no 
&&L_{\rm 2-loop}^{(c)}=
\hbar \mu\left[8 a_{_b}^4(n+1)(n+2)+8 (-a_{_b})^4(n+1)(n+2)\right]I_2(\varepsilon)\ 
\STr[g^{-1}\Psi_{_L} g\Psi_{_R}]\
\cr
&& \ ~~~~~~~ \  =\hbar \mu\left[2 a_{_b}^4(n+1)(n+2)+2 (-a_{_b})^4(n+1)(n+2)\right]\ 
[I_1(\varepsilon)]^2\  \STr[g^{-1}\Psi_{_L} g\Psi_{_R}]\ , 
\nonumber
\ee
where the integrals $I_1(\varepsilon)$ and $I_2(\varepsilon)$ 
were  defined in eqs. (\ref{ii1}) and (\ref{i2}), respectively, 
and in the last line   we used eq. \rf{onr}  relating $I_2$ and $(I_1)^2$.

It is interesting to note that each one of the above three
contributions is proportional to $(n+1)(n+2)$. This factor may be
understood on the group theory grounds as being the product of the two
quadratic Casimirs, in the fundamental and the adjoint representations
of $Sp(n-2,2)$ or $Sp(n)$.
%
 This $n$ dependence is different from that of the corrections to the
 bosonic potential because, on the one hand, 
 in the bosonic calculation one uses that (see \rf{ip}) 
 $T^2=- {1 \ov 4} \id$ while here the analogous
quantities are  $\Psi_{_L}^2$ or $\Psi_{_L}\Psi_{_R}$ do not
 have similar properties, and, on the other hand,
some Wick contractions  here are  forbidden as the fields belong
to different algebras.

Adding together the above three singular contributions in (\ref{fermi_b})
 we conclude, in complete analogy with the bosonic potential case,
 that they cancel out, i.e. the result is UV finite,
\be
L_{\rm 2-loop}^{(\rm fermi.pot.)}={\rm finite } \ . 
%
\ee

\section{Concluding remarks}

The reduced model \rf{lag} \ci{gt,ms} we discussed above is naturally
associated, through the Pohlmeyer reduction, to the \adss GS
superstring action \rf{al} and has certain unique features.
 
Its construction is based on first-order or phase space formulation of
superstring dynamics in terms of supercoset currents,  with the Virasoro
constraints explicitly solved in terms of a new set of variables related
locally to currents and thus non-locally to the original GS \adss
supercoset coordinates.  Although various steps in the reduction do
not appear to manifestly preserve 2d Lorentz invariance, 
the resulting reduced Lagrangian describes the dynamics of the
physical number of  degrees of freedom in a manifestly
Lorentz invariant way. Being formulated in terms of left-invariant
currents, the reduced theory is apparently ``blind'' to the original
global $PSU(2,2|4)$ symmetry; however, being integrable (the Lax pairs
of the original
and the reduced theory are gauge-equivalent), it still has an
infinite number of commuting charges associated to hidden symmetries,
some of which are  implicitly 
 related to the global symmetries of the
original GS theory.

In general, the Pohlmeyer reduction procedure, utilizing the classical
conformal symmetry of a 2d sigma model, is expected to lead to an
equivalent theory only at the classical level; for example, the
original and reduced theory are obviously not equivalent at the
quantum level if the original sigma model has a  running coupling.
In the present case of \adss superstring sigma model, which is  
a conformal 2d theory at the quantum level, the relation between the
original and the reduced theory has a perfect chance to hold also at
the quantum level. The  necessary condition for that is 
 that the reduced theory is
also UV finite.

 As we have demonstrated in the present paper, the reduced theory
 associated to the \adss superstring model is indeed free of 2d UV
 divergences in a certain renormalization scheme.  
 An advantage of the reduced theory compared to the GS
 model is that here the main ``kinetic'' part of the action is based on a 
 gauged WZW theory  and thus is guaranteed to be finite;
 then   what remains to
 check is only the absence of divergent contributions to the
 derivative-independent ``potential'' part of the action. We
 explicitly checked that  at the 1-loop and 2-loop order but most likely
 this should be true to all orders and should be due to a hidden 2d
 (super)symmetry of the reduced theory.\foot{If the reduced theory
 does not actually have a standard global 2d supersymmetry, this
 finiteness property suggests that there may be other similar models
 without 2d supersymmetry that are still UV finite. It would be
 interesting to classify them.}  The cancellation of divergences is
 due to a very special balance between the bosonic potential term
 and the fermionic interaction term in \rf{lag}.  These two terms
 originated from the ``kinetic'' $P^2$ and the fermionic WZ $Q^2$
 terms in the GS action \rf{al} where they were related by
 $\k$-symmetry. This suggests that some (global) remnant of the
 $\k$-symmetry   still present after  fixing the 
 $\k$-symmetry gauge
  in the reduced action  may be responsible for its UV finiteness.

This opens up a possibility of solving 
the quantum \adss superstring theory in terms of the
the quantum reduced theory. The precise prescription for 
translating observables between the two theories
remains to be understood. The most optimistic scenario is to find a
path integral version of the reduction procedure based on changing the
variables from coordinates to currents and solving the conformal gauge
constraints as delta-function conditions $T_{++}=0, \ T_{--}=0$ in the
path integral.

To test the equivalence of the two partition functions one may
consider comparing their values for 
equivalent classical solutions. We leave the study of this problem for
the future.  
Among other open problems let us mention the construction of the (2d
Lorentz-invariant) S-matrix for  scattering of the massive 
elementary excitations in the reduced theory
and 
the determination of
its relation to the BMN (magnon) S-matrix in the \adss
string theory in a light-cone gauge.

 Let us finish with few comments on the role of the $\mu$ parameter in
 the reduced theory. The original GS string theory 
 in conformal gauge has a residual part of the 2d diffeomorphism group
 -- conformal reparametrizations -- being preserved by quantum
 corrections. In the process of constructing the reduced theory we fix
 this residual symmetry by a gauge choice (cf. \rf{ret}) that
 introduces the constant parameter $\mu$. This parameter is a fiducial
 scale, similar to the constant $p^+$ in the standard light-cone
 gauge.\foot{Indeed, the condition $P_+ = \mu T$ in \rf{ret} is
 reminiscent of the relation $ \del_+ x^+ \sim p^+$ in the light-cone
 gauge.  Compared to standard 2d conformal theories where the
 infinite-dimensional conformal group is interpreted as a global
 symmetry imposed through conditions on physical states, in the
 context of string theory this is part of the 2d diffeomorphism gauge
 symmetry and one is allowed to fix it by a gauge choice.} Thus $\mu$ is
 similar to a gauge-fixing parameter and physical observables should
 not depend on it.  For example, the expression for the energy of a
 particular string state expressed in terms of conserved charges of
 the reduced theory (or, e.g., Casimirs of the original GS global
 symmetry group) should not depend on $\mu$, i.e. $\mu$ can be
 eliminated by re-expressing it in terms of the charges.  At the same
 time, the $S$-matrix of elementary excitations with mass $\mu$
 (which, by itself, is not a physical observable) will depend on
 $\mu$.\foot{One may draw an analogy with quantization of strings in
 plane wave background. In conformal gauge one has a sigma model with
 target space metric like 
 $ds^2=dx^+ dx^- + a x_i x_i dx^+ dx^+ +
 dx_i dx_i$ and certain global symmetry group. One may, in principle,
 develop a covariant quantization and find the spectrum of states
 which will be classified by charges of that symmetry. We may instead
 fix the light-cone gauge $x^+ = p^+ \tau$ and obtain a model containing
 free bosons (and fermions, as in the pp-wave model \ci{mets,mbnm,bmn}
 associated to \adss background) with mass $\mu= p^+$.  Then the spectrum will
 depend on that $\mu$, but we may re-interpret that dependence as
 that  on one of the global charges which  has a fixed value
 (proportional to $\mu$) in that light-cone gauge.}

\section*{Acknowledgments }

We are grateful to M. Grigoriev, R. Metsaev,  G. Papadopoulos  and A. Vainshtein for useful
discussions. AAT acknowledges the support of the STFC rolling grant.
Part of this work was done while AAT was a participant of the 2008
workshop ``Non-Perturbative Methods in Strongly Coupled Gauge
Theories''
at the Galileo Galilei Institute for Theoretical Physics in 
Arcetri, Florence.
RR acknowledges the support of the US National Science Foundation under
grant PHY-0608114, the US Department of Energy under contract
DE-FG02-90ER40577 (OJI) and of the A.~P. Sloan Foundation.


\

\appendix
\subsection*{Appendix A:  Comments on regularization scheme ambiguity
}
\refstepcounter{section}
\def\theequation{A.\arabic{equation}}
\setcounter{equation}{0}

%

Regularization scheme dependence of the 2-loop corrections to the
bosonic and fermionic potentials implies that while apparently
different results may be obtained under different choices of
regularization (and, in particular, of treatment of fermions and
Levi-Civita tensors), all of them are related by suitable
redefinitions of the coupling constants of the theory.  The most
natural regularization scheme should be consistent with the symmetries
of the theory, and we believe the dimensional reduction regularization
used in the main text is such a scheme, though that seems non-trivial
to demonstrate explicitly.\foot{ An intuitive reason is that the
reduced model is related to the \adss GS superstring where the
$\kappa$-symmetry should be preserved. The 2-loop finiteness of the
\adss superstring demonstrated in \cite{rt} in this scheme is a strong
indication in this direction.}  For completeness, in this Appendix we
discuss the 2-loop results in some alternative regularization schemes.

A version of dimensional regularization prescription (which {\it does
not}, however, preserve the $d$-dimensional Lorentz invariance) is to
continue momenta to $d=2-2\vep$ from the very beginning while still
treating the Levi-Civita tensor $ \epsilon^{\mu\nu}$ as if it is
defined only in 2-dimensions \ci{bos} (i.e. $\eps^{\mu\nu} \to \bar
\eps^{\mu\nu}\equiv \eps^{\bar \mu\bar \nu}, \ \ \bar \mu,\bar \nu=1,2
$). Then instead of \rf{eeee} we get 
\bea
&&-(\bar \epsilon^{ \mu \nu}p_\mu q_\nu)^2
={\bar p}^2 {\bar q}^2-({\bar p}\cdot{\bar q})^2
=[(p^2-{\hat p}^2) (q^2-{\hat q}^2)-(p\cdot q-{\hat p}\cdot{\hat q})^2] \no \\
&&=[p^2q^2-(p\cdot q)^2] -\left[p^2{\hat q}^2+q^2{\hat p}^2
-2p\cdot q\,{\hat p}\cdot{\hat q}\right]+
[{\hat p}^2 {\hat q}^2-({\hat p}\cdot{\hat q})^2] \ . \la{kii}
\eea
Here ${\bar p}$ and ${\hat p}$ are the $2$-dimensional and
$-2\vep$-dimensional components of the momentum $p$ in $d=2-2\vep$
dimensions, $p^\m= (p^{\bar \mu}, p^{\hat \mu} )$.  The contribution
of the first square bracket in the last line to the integral in
\rf{i2} is then the same as in
\rf{onr}, while the second bracket leads to
\be
&& 
- \int\frac{d^dp d^d
q}{(2\pi)^{2d}}\frac{\left[p^2{\hat q}^2+q^2{\hat p}^2
-2p\cdot q\,{\hat p}\cdot{\hat q}\right]}{p^2 q^2 [(p+q)^2]^2}=-
\int\frac{d^dp d^d q}{(2\pi)^{2d}}
\frac{{\hat p}\cdot{\hat q}}{p^2 q^2 (p+q)^2}  \cr 
&&
=
-\frac{{\hat \eta}_{\mu\nu}\eta^{\mu\nu}}{d}
\int\frac{d^dp d^d q}{(2\pi)^{2d}} \frac{{p}\cdot{q}}{p^2 q^2 (p+q)^2}
=-\frac{\epsilon}{d}[I_1(\vep)]^2 \ ,  \la{rkj}
\ee
%
where we used that for $d=2-2\vep$ one has ${\hat
\eta}_{\mu\nu}\eta^{\mu\nu} = - 2 \vep$. The integral of the remaining
square bracket in the last line of \rf{kii} may be written as
\be
({\hat \eta}_{\mu\nu}{\hat \eta}_{\rho\sigma}-
{\hat \eta}_{\mu\rho}{\hat \eta}_{\nu\sigma})
\int\frac{d^dp d^d q}{(2\pi)^{2d}}\frac{p^\mu p^\nu q^\rho q^\sigma}
{p^2 q^2 [(p+q)^2]^2} \ ,   \la{jq}
\ee
and 
produces a finite $O(\vep^2) [I_1(\vep)]^2 $ contribution.  As a
result, the expression for $I_2$ in  \rf{i2} in this regularization scheme is
given by the sum of \rf{onr},\rf{rkj} and \rf{jq}, i.e.
\be 
I_2
=  \Big[{1 \ov 4}  - {\vep \ov 2}  +   {\cal O}(\vep^2)\Big]  [I_1(\vep)]^2 
\la{hio} \ .  
\ee
The contribution of the  diagram  (c) in (\ref{figc}) is then 
\be 
\la{ccc_A}
 L^{(c)}_{\rm2-loop}=
 4\hbar \mu^2 a_{_b}^4 \  n(n+2) \  \Big[ 
 1 - 2  \rb \vep +   {\cal O}(\vep^2)\Big]  [I_1(\vep)]^2       \
\STr[g^{-1}TgT]  \ , 
\ee 
where $\rb=0$ in the dimensional reduction regularization 
used  in section \ref{sec:bose_pot} 
with $I_2$  given by \rf{onr} and $\rb=1$ in the second regularization
prescription where $I_2$ is given by \rf{hio}.
The total bosonic contribution is then 
\be
L^{\rm bose}_{\rm 2-loop} = \frac{\hbar \m^2}{32}\ 
\left[n^2 -    \rb   n(n+2)  {\vep}+{\cal O}(\vep^2)\right] \ [I_1(\vep)]^2\  \STr[g^{-1}TgT] \ , 
\label{bose22}
\ee
where $\rb=0$  corresponds  to \rf{bose2}.


Thus, unlike what happened in the regularization by dimensional reduction, 
the bosonic contribution to the 2-loop anomalous dimension 
does not vanish in this ($d$-dimensional Lorenz-violating) scheme. 
The resulting   value for the 2-loop anomalous dimension 
%
%
is, however,  in agreement with the standard expression for the two-loop
anomalous dimension in a sigma model with a WZ coupling (see
discussion below eq.\rf{beel}) and, in particular, with the expression
for the anomalous dimension of the primary field $\tr g$ in WZW theory
\ci{kz} in \rf{cec}.\foot{In the bosonic theory  with the  group  $Sp(n-2,2)
\times Sp(n) $ we have  the kinetic and potential terms for each factor
 decoupled,  so that  for, e.g., $G=Sp(n)$ 
 we get  for the two anomalous dimensions, cf.\rf{zez},\rf{zezz} 
($c_{_G}= c_{_{Sp(n)}} = n+2$)

 $
 \g (Sp(n-2,2)) =  {c_{1} \ov k} ( -1  +  { c_{_G}\ov 2 k } + ... )  \ , 
\ \ \ \ \ \ \ \ \ \ \ 
  \g (Sp(n)) =  {c_{1} \ov k} ( 1   +   {  c_{_G}\ov 2 k } + ... )  \ ,
   $  
%

\noindent
where $c_1= c_{_r}=n+1$ (=Casimir of the fundamental representation of $Sp(n)$) 
 in the case of the $\tr g$ operator  and $c_{1}=n$
in the present case of the  $\tr( g^{-1} T g T ) $ operator 
(cf.  \ci{jjp,om}; for comparison,  in the case of  $\tr( g^{-1} T^a g T^b ) $
where   $T^a$ are generators of $G$  one has $c_{1}= c_{_G} $ \ci{kz}).
Going from one group factor to another is thus equivalent to $k \to
-k$ (notice that we had $\STr$ in the WZW kinetic term in \rf{lag} and
\rf{edS}).
}

\

Similarly to the treatment of $I_2$ there are  several options of how to define 
 the integral $I_3$ \rf{tri}, i.e. of how to extend  it to $d$ dimensions.
%
Instead of using the dimensional reduction scheme we 
 may choose  to  extend momenta to $d$ dimensions  from the start but 
 treat  the indices  of the  integrand factor $p_+ q_-$   in \rf{tri}
 as 2-dimensional ones. Then instead of \rf{ttq}  we have  ($\bar \m, \bar \n=1,2$) 
 \be 
\la{ttqs} 
p_+ q_- = (p_0 + p_1) (q_0-q_1) = -\eta^{\bar\m\bar \n} p_{\bar \m}  q_{\bar \nu } -   \eps^{\bar\m\bar \n} p_{\bar \m}
 q_{\bar \nu }
 \ ,  
\ee 
and computing the integral  in \rf{tri}  gives,   instead of \rf{irs},\foot{One more option is
to use the straightforward dimensional regularization where $ \langle
p_\m q_\n \rangle = { 1 \ov d} \eta_{\m\n} \langle p \cdot q\rangle $
and thus  $ \langle p_+ q_- \rangle = - { 2 \ov d} \langle p \cdot
q\rangle$. In this case  $ I_3 ={ 1 \ov d}[I_1(\vep)]^2$ leading to 
$\ha (1 + 2 \vep) { 1 \ov (4 \pi)^2 \vep^2}$ divergent term.} 
\be \la{yyy}
I_3= - \int {d^2 p\ov (2 \pi)^2 }{  d^2 q\ov (2 \pi)^2 } \frac{   (\bar p\cdot  \bar q) }
{p^2q^2(p+q)^2} =  {1 \ov 2} \big( 1 +  { 2 \vep \ov d} \big)  [I_1(\vep)]^2 \ . 
\eea
Then 
\be\la{hoi}
L^{\rm fermi}_{\rm2-loop}=  - \frac{\hbar \mu^2}{32}  \ 
n^2 (1 + {\rm f}\vep ) \   [I_1(\vep)]^2 \ \Str[g^{-1}Tg T]  \ , 
\ee
where ${\rm f}=0$  corresponds to the dimensional reduction prescription 
used in \rf{hopi} and ${\rm f}=1$ corresponds to the above 
prescription leading to \rf{yyy}.

Combining this with the bosonic contribution in
\rf{bose22} we conclude that the leading $ 1 \ov \vep^2$ singularity
cancels out between the bosonic and the fermionic terms, just as the
corresponding $ 1 \ov \vep$ singularity did at one loop, and  we are left  with
\be
L_{\rm 2-loop}^{(\rm bos.pot.)}=
L^{\rm bose}_{\rm 2-loop}+L^{\rm fermi}_{\rm 2-loop}&=& 
-  \frac{\hbar\m^2}{32} 
\Big[ [\rb n (n+2)  + {\rm f} n^2] \vep +  {\cal O}(\vep^2)\Big] 
\,[I_1(\vep)]^2\,\STr[g^{-1}TgT]~~\cr
&=&
-  \frac{\hbar\mu^2}{32 (4\pi)^2  \vep}  [\rb n (n+2)  + {\rm f} n^2]
\  \STr[g^{-1}TgT]+{\rm finite}
 \ . \la{pqw_A} 
\ee 
This remaining divergent term is clearly regularization-scheme
dependent and may be set to zero by an appropriate {\it finite}
redefinition of the couplings (in particular, the level of the WZW model).

\

\appendix
\subsection*{Appendix B:  Analogy with 2d supersymmetric 
sigma models with potentials}
\refstepcounter{section}
\def\theequation{B.\arabic{equation}}
\setcounter{equation}{0}

\def \WW {{\cal W}} 

It is important to note that the dimensional reduction scheme in which
the reduced model is 2-loop finite is also the scheme that would
preserve 2d supersymmetry, if it were present at the classical level.

It is useful to draw analogy with a general analysis of 2-loop
renormalization of $(p,q)$ supersymmetric models deformed by
potentials \ci{pta} carried out in \cite{lam}.  A special case of the
model considered in \ci{lam} is the (1,1) supersymmetric theory
generalizing a supersymmetric WZW model to the presence of a potential
term \ci{pta} (cf. \rf{sip})
\bea 
&&S= { 1 \ov 4 \pi \a'} \int d^2 \s \Big[ 
(G_{mn}(x) + B_{mn} (x)) \del_+ x^m \del_- x^n  
+ i  G_{mn}(x) \psi^m_{_L} D^{(+)}_+ \psi^n_{_L}
+ i  G_{mn}(x) \psi^m_{_R} D^{(-)}_- \psi^n_{_R}\cr 
&& \ \ \ \ \ \ \ \ \ \ \ \ 
+\  2 \mu  D^{(-)}_m W_n(x) \psi^m_{_L}\psi^n_{_R} - \m^2  G^{mn}(x) W_m (x) W_n (x) \Big]
 \ . 
 \la{sus}
 \eea
Here $G_{mn}$ and $B_{mn}$ correspond to a group space $G$, $x^m$ are
coordinates on $G$, $D^{(\pm )}$ are covariant derivatives with
respect to the two ``flat'' connections $ \Gamma^m_{nk} (G) \pm \ha
H^m_{\ nk}(B)$,\foot{ As is well known, the kinetic terms of the
fermions can be decoupled from bosons by defining the tangent space
components like $\psi^a= E^a_m(x) \psi^m$ and ``rotating'' $\psi^a$.} 
and a vector $W_m$ defines the bosonic potential.
 
In general \ci{pta}, $W_m = U_m - V_m$, where $ D_{(m} V_{n)} =0$
(i.e. $V_m$ is a Killing vector), $ \del_{[m} U_{n]} =\ha H_{mnk}
V^k$, \ $U_m V^m = 0$.  The condition of 1-loop (and, in fact, 2-loop)
finiteness of such model is \ci{lam}\ $D_m W^m=\const$.
  
In the simplest case $W_m = \del_m {\WW}$ where $\WW$ is real (1,1)
superpotential.  In that case the action \rf{sus} can be written in
the superfield form: 
\be 
S= { 1 \ov 4 \pi \a'} \int d^2 \s d^2 \theta
\Big[ (G_{mn}(X) + B_{mn} (X)) \hat D_+ X^m \hat D_- X^n - \WW(X)
\Big] \ , \la{susa}
\ee
where $X^m = x^m + \theta_+ \psi^m_{_L} + 
 \theta_- \psi^m_{_R}  + \theta_+ \theta_- F^m$ and $\hat D$ are 
spinor derivatives.\foot{
%
%
The (1,1) supersymmetric WZW action can also be written explicitly in
terms of a superfield generalizing the group element $g$ field
\ci{dik}.
 Explicitly, we may replace $g= e^x$ by $\hat
g= e^{X}$, \ $X(\s,\theta) = x + \theta_+ \psi_{_L} + \theta_-
\psi_{_R} + \theta_+ \theta_- F$. Then to supersymmetrize the
potential $\tr( g^{-1} T g T)$ we need to find the corresponding real
superpotential $\WW$. 
This step is  straightforward for coset sigma models of the type
\rf{hgi} whose potential depends on only two special fields $\varphi$ 
and $\phi$ such that the (1,1) superpotential may be written as $\cosh
\hat \phi + \cos \hat \varphi $ or as 
Re$[\cos (\hat \varphi + i \hat \phi )]$. 
%
%
Note that the holomorphic superpotential of the (2,2) sine-Gordon
model found \ci{gt} in the special case of the model
\rf{sinh} is ${\rm W}= \cos( \hat \varphi + i \hat \phi )$, but more general
models like \rf{hgi} do not admit a straightforward (2,2) extension as
$\varphi$ and $\phi $ enter separately in the two factors of the
target space metric. }

In the 2d theory \rf{sus} the bosonic and the fermionic potential
terms renormalize simultaneously, i.e.  the $\beta$-functions of the
corresponding couplings are related by a supersymmetry Ward identity.
As was shown in
\cite{lam}, the 2-loop correction   to this 
$\beta$-function vanishes in the dimensional reduction scheme similar
to the one used here in section \ref{sec:bose_pot}.  Thus in the (1,1)
supersymmetric theory \rf{sus} and the reduced theory \rf{edS} both
treated in the dimensional reduction scheme there are no genuine
2-loop simple-pole UV divergences, all of them being accompanied by a
double-pole counterpart related to single-pole 1-loop divergences as
dictated by the renormalizability of the theory.

The model \rf{edS} based on $G=G_1 \times G_2$ bosonic WZW model with
a potential coupled to fermions in bi-fundamental representations does
not admit the standard version of (1,1) 2d supersymmetry: the standard
supersymmetric extension of its bosonic part would be of the form
\rf{susa}, i.e.  having the same number of the fermionic degrees of
freedom but transforming in the adjoint representation of $G$. The
corresponding $G_1$ and $G_2$ supersymmetric models would 
be mutually non-interacting
and the divergences in their potential terms will not cancel,
precluding finiteness.

The non-trivial property of the reduced model observed here is the
cancellation of the 1-loop divergences, 
which makes the theory (at least) 2-loop finite. Such finiteness
property is also characteristic of $(2,2)$ supersymmetric models
\ci{lam}.  The existence of a finite $(2,2)$ supersymmetric extension of
a bosonic WZW model (with a group $G$ which is a complex manifold)
perturbed by a potential appears to be subtle and we are not aware of
its discussion in the literature.\foot{The existence of a (2,2) 
superpotential deformation for the supersymmetric WZW models discussed
in this appendix is, to some extent, questionable. Indeed, the
relevant superpotential should be a holomorphic function on the target
space. However, the target space here is factorized with one factor
being compact, implying that any holomorphic function on this part of
the target space is constant. More general approaches to the
construction of supersymmetric extensions of sigma models with torsion
encounter difficulties due to the rather trivial topology of
semi-simple groups. 
We thank G. Papadoupoulos for useful comments
on these issues. 
%
} 





\end{document}